\documentclass[aps,prb,preprint,floatfix]{revtex4}

\usepackage{amsmath}  
\usepackage{amsfonts} 
\usepackage{graphicx} 

\begin{document}

\title{The Electric Field of a Uniformly Charged Non-Conducting Cubic
  Surface}

\author{Kaitlin McCreery}
\email{kpmccreery@gmail.com}
\author{Henry Greenside}
\email{hsg@phy.duke.edu}
\affiliation{Department of Physics, Duke University, Durham
NC 27708-0305}

\date{\today}

\begin{abstract}
  As an integrative and insightful example for undergraduates learning
  about electrostatics, we discuss how to use symmetry, Coulomb's law,
  superposition, Gauss's law, and visualization to understand the
  electric field~${\bf E}(x,y,z)$ produced by a non-conducting cubic
  surface that is covered with a uniform surface charge density. We
  first discuss how to deduce qualitatively, using only elementary
  physics, the surprising fact that the electric field inside the
  cubic surface is nonzero and has a complex structure, pointing
  inwards towards the cube center from the midface of each cube and
  pointing outwards towards each edge and corner. We then discuss how
  to understand the quantitative features of the electric field by
  plotting an analytical expression for~${\bf E}$ along symmetry lines
  and on symmetry surfaces. This example would be a good choice for
  group problem solving in a recitation or flipped classroom.
\end{abstract}

\maketitle

\section{Introduction}
\label{sec:intro}

An undergraduate learning about electrostatics has to learn many
challenging concepts such as symmetry (of a charge distribution or of
a field), a three-dimensional electric vector field~${\bf E}(x,y,z)$,
a three-dimensional electric potential scalar field~$V(x,y,z)$, and
related ideas such as superposition, Coulomb's law, and Gauss's law.
Introductory\cite{Tipler2007,Serway2011,Young2011,Knight2012,Cutnell2014}
and upper-level\cite{Griffiths2012,Purcell2013} textbooks explain
these concepts and provide examples that illustrate how to solve
problems related to these concepts, but many physics curricula, some
of which cover material at the rapid rate of one textbook chapter per
week, lack the time to give students sufficient practice to master
these concepts. Especially lacking are integrative examples that use
multiple concepts (say spanning several chapters of a textbook) and
that use different problem solving strategies (say qualitative,
analytical, numerical, and visual).

\begin{figure}
\centering
\includegraphics[width=0.3\textwidth]{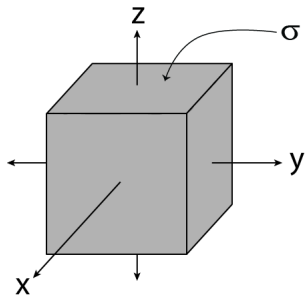}
\caption{A three-dimensional charge distribution that consists of a
  uniform positive surface charge density~$\sigma$ on a non-conducting
  cubic surface. The center of the cube coincides with the origin~$O$
  of the $xyz$~Cartesian coordinate system used in this paper. The
  cube has sides of length~$L=1\,\rm m$ so that the eight vertices lie
  at the coordinates $(x,y,z) = (\pm 1/2, \pm 1/2, \pm 1/2)$ in
  meters.}
\label{fig:charged-cubic-surface}
\end{figure}

To help students gain a deeper understanding of electrostatics and to
improve their qualitative and quantitative problem solving skills, we
discuss an electrostatic problem that is conceptually and technically
appropriate for students taking an introductory physics course or an
upper-level course on electricity and magnetism. The problem concerns
understanding the electric field produced by a symmetric
three-dimensional charge distribution that consists of a
non-conducting cubic surface that is covered with a uniform positive
charge density~$\sigma$ (see Fig.~\ref{fig:charged-cubic-surface}).

This example is insightful in several ways. First, it shows students
how to use qualitative reasoning based on symmetry, superposition,
Coulomb's law, and Gauss's law to deduce the properties of an electric
field that is substantially more complicated than what most
undergraduate physics courses discuss.  For example, we explain how to
deduce qualitatively that the electric field inside the charged
non-conducting cubic surface points inwards near the middle of cube
faces and points outwards towards the edges and vertices. The fact
that the electric field inside this surface is nonzero by itself will
likely be surprising to students because this example seems similar to
other introductory physics examples for which the electric field is
zero inside a symmetric charge distribution. (One example is the
electric field inside a spherical surface covered with a uniform
surface charge density, while a second example is the field inside a
charged conducting cubic surface, see
Section~\ref{subsect-related-systems-with-zero-field} below.)  That
many interesting insights can be deduced qualitatively using
elementary physics is one of the main contributions of this paper.

It turns out that one can obtain an explicit, although long, analytic
expression for the potential~$V(x,y,z)$ everywhere in space for a
uniformly charged cubic surface (see
Appendix~\ref{appendix:exact-electric-field} and
Ref.~\onlinecite{Trott2012}). Via the relation~${\bf E} = - \nabla V$,
the expression for~$V$ leads to an even longer explicit expression for
the electric field. We use the analytical expression for~${\bf E}$ to
confirm and then to extend the insights obtained by qualitative
reasoning. Because the expression for~${\bf E}$ is too long to provide
insight, we demonstrate the value of plotting the analytical
expression for~${\bf E}$ on certain lines and areas that have
symmetries such that the electric field is everywhere parallel to
these lines and areas.  Continuity of the electric field then allows
one to extend knowledge of the field on the symmetry lines and
surfaces to the full three-dimensional interior, and so provides a
rather complete understanding of the interior electric field.

We envision this paper being used in several ways by physics
instructors. Simplest is to use parts of this paper as supplementary
homework problems or to stimulate thinking and discussion in class by
using parts of this paper for interactive questions, say via anonymous
polling\cite{Weiman2005,Caldwell2006}. But we feel that this paper
will be most useful as a group learning project in a recitation or
flipped class, in which students in small groups collaborate to work
through portions of this paper in guided steps.

The rest of this paper is organized as follows. In
Section~\ref{sec:qualitative}, we explain how to use qualitative
reasoning to deduce key properties of the electric field inside the
uniformly charged cubic surface. This section is divided into
subsections that represent pedagogical milestones that a group of
students could discuss and solve in succession. In
Section~\ref{sec:quantitative}, we use an analytical expression for
the electric field of this cubic surface to confirm the qualitative
arguments of Section~\ref{sec:qualitative} and then discuss some new
quantitative features of the electric field and of the potential~$V$.
In Section~\ref{sec:conclusions}, we summarize our results and discuss
some of their pedagogical
implications. Appendix~\ref{appendix:exact-electric-field} discusses
the analytical expression for the potential and electric field of the
uniformly charged cubic surface, and uses the analytical expression
for~${\bf E}$ to show that it diverges at the edges of the cubic
surface.

\section{Qualitative Insights}
\label{sec:qualitative}

In this section, we use qualitative reasoning based on symmetry,
superposition, continuity, Coulomb's law, and Gauss's law to deduce
the properties of the electric field~${\bf E}$ produced by a
non-conducting cubic surface that is covered with a uniform positive
surface charge density~$\sigma > 0$. A surprising conclusion is that
the interior electric field is not zero and has a complicated
geometry.  Section~\ref{sec:quantitative} then uses an analytical
expression for~${\bf E}$ to confirm and to extend these qualitative
insights by providing quantitative details.

\subsection{Only the interior electric field  is interesting to consider}
\label{subsection-only-interior-field-interesting}

A first point is that only the electric field interior to the
uniformly charged non-conducting cubic surface is interesting to
explore. This is because the electric field exterior to the cubic
surface is qualitatively similar to the familiar electric field of a
positive point charge at the center of the cubic surface: ${\bf E}$
points away from the center of the cubic surface, and it diminishes in
magnitude with increasing distance from the center of the cube.

One way to see this is to observe that, for any point~$P$ outside the
cubic surface, one can find a plane such that the point~$P$ is on one
side of the plane and the entire cubic surface is on the other side.
(For example, the line segment connecting~$P$ and the center~$O$ of
the cubic surface must intersect a face of the cube. Any plane outside
the cubic surface that is parallel to that face and that intersects
the interior part of the line segment~$PO$ will suffice.)  Since the
infinitesimal point charges that make up the cubic surface then lie on
one side of the plane, the total electric field~${\bf E}$ at point~$P$
due to these point charges must point away from the cubic surface
(although generally not radially, i.e., not parallel to the line
segment $PO$).

In contrast, the electric field interior to the cubic surface can be
complicated precisely because, at some point~$P$ inside the cubic
surface, there are contributions to the electric field at~$P$ from all
possible directions, associated with the infinitesimal point charges
making up the surface charge density. However, as we now discuss,
these contributions generally do not cancel to give zero.

\subsection{Electric field vectors are parallel to mirror 
planes at points on such planes}
\label{subsection-E-parallel-to-mirror-planes}

A next step in our qualitative analysis is to take advantage of the
symmetries of the uniformly charged cubic surface. These symmetries
provide a way to deduce quickly and without calculation some
information about the direction of the electric field on certain
planes and lines, which are then the locations to consider first when
trying to understand the electric field. In
Section~\ref{sec:quantitative}, we will see that these symmetry planes
and lines are also good places to plot quantitative information about
the electric field.

\begin{figure}
\centering
\includegraphics[width=0.5\textwidth]{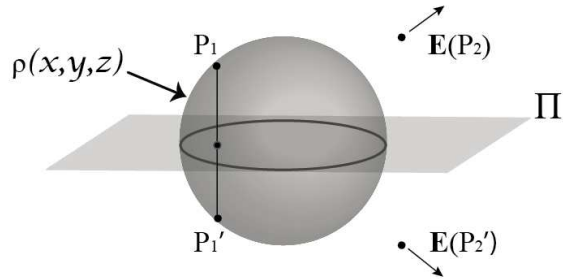}
\caption{A charge distribution~$\rho$ (here a sphere filled with a
  spherically symmetric charge density~$\rho(r)$) has a mirror
  symmetry plane~$\Pi$ if the plane divides the charge distribution
  into two distinct sets such that, for every point~$P_1$ in one set,
  there is a corresponding point~$P_1'$ in the other set such
  that~$P_1$ and~$P_1'$ are mirror images with respect to~$\Pi$.  The
  mirror symmetry of~$\rho$ implies that electric field~${\bf E}$
  created by~$\rho$ has a mirror symmetry. This means that the
  electric field vector~${\bf E}(P_2)$ at some point~$P_2$ and the
  electric field vector~${\bf E}(P_2')$ at the mirror image
  point~$P_2'$ are themselves mirror images of each other as shown. As
  the points~$P_2$ and~$P_2'$ approach the symmetry plane~$\Pi$, the
  corresponding electric field vectors become identical and parallel
  to~$\Pi$.}
\label{fig:mirror-plane-diagram}
\end{figure}

A charge distribution~$\rho(x,y,z)$ is said to have a mirror symmetry
plane (or mirror plane for short) if there is a plane~$\Pi$ that
divides the charge distribution into two distinct parts that are
mirror reflections of each other with respect to~$\Pi$ (see
Fig.~\ref{fig:mirror-plane-diagram}). This means that for every
point~$P_1$ of the charge distribution that lies on one side of the
plane~$\Pi$, there is a corresponding point~$P_1'$ (the mirror image
of~$P_1$ with respect to plane~$\Pi$) on the other side of the plane
such that the plane is perpendicular to and bisects the line
segment~$P_1P_1'$. The points~$P_1$ and~$P_1'$ are related in the same
way that one finds the image~$P_1'$ of a point~$P_1$ with respect to a
planar mirror via geometric optics\cite{Knight2012}.

\begin{figure}
\centering
\includegraphics[width=0.9\textwidth]{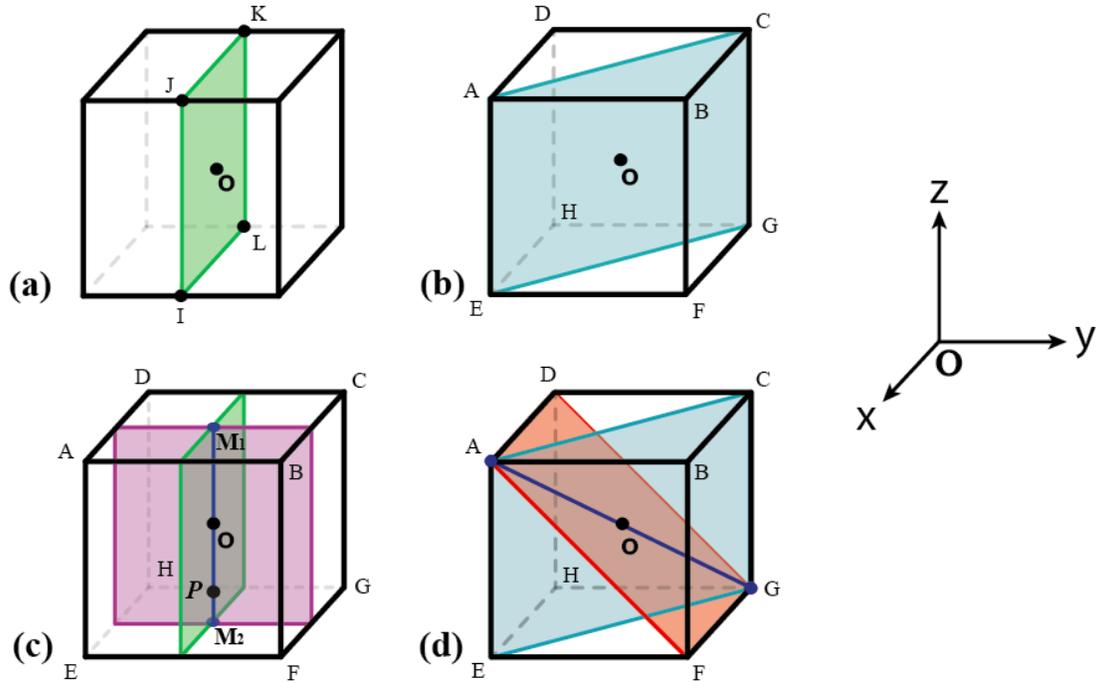}
\caption{(a) The square~IJKL lies in one of three mirror planes of the
  cube that passes through the cube's center~$O$ and through the
  midpoints of four parallel edges. (b) The rectangle~ACGE lies in one
  of six mirror planes that pass through the cube's center~O and
  through two pairs of diagonally opposite corners. (c) The line
  segment~$M_1OM_2$ is one of three symmetry line segments that pass
  through the cube's center and that connects the midpoints~$M_1$
  and~$M_2$ of two opposing faces. 
  (d) The line segment~$AOG$ is one of four symmetry line segments
  that passes through~$O$ and through two diagonally opposite corners,
  here~$A$ and~$G$. 
 }
\label{fig:symmetry-planes-and-lines-of-cubic-surface}
\end{figure}

Most charge distributions discussed in introductory physics courses
have mirror planes, for example a point charge, a uniformly charged
line segment, a cylindrical surface filled with an azimuthally
symmetric charge density, a sphere filled with a spherically symmetric
charge density, and an infinite uniformly charged plane. As shown in
Fig.~\ref{fig:symmetry-planes-and-lines-of-cubic-surface}, a uniformly
charged cubic surface has nine distinct mirror planes. Three of these
planes (Fig.~\ref{fig:symmetry-planes-and-lines-of-cubic-surface}(a))
have the property of passing through the center~$O$ of the cube and of
passing through the midpoints of four parallel edges.  The six other
mirror planes
(Fig.~\ref{fig:symmetry-planes-and-lines-of-cubic-surface}(b)) have
the property of passing through the center of the cube and of passing
through two pairs of diagonally opposite vertices.

A mirror plane~$\Pi$ of a charge distribution is useful because, at
any point~$P$ on such a plane, the electric field vector~${\bf E}$
at~$P$ is parallel to~$\Pi$. This follows from the experimentally
observed uniqueness of motions governed by Newton's second law $m {\bf
  a} = {\bf F}$ when applied to a point particle of mass~$m$ and
charge~$q$ in an electric field so that the force is given by~${\bf F}
= q {\bf E}$. If a charged point particle is placed at some
point~$P_2$ in some electric field with zero initial velocity (see
Fig.~\ref{fig:mirror-plane-diagram}), the particle is always
observed to move along a unique path, which implies that there can
only be one force vector at~$P_2$ and so only one electric field
vector~${\bf E} = {\bf F}/q$. It then follows that the electric field
at any point on a mirror plane cannot have a component perpendicular
to the plane (i.e., ${\bf E}$ must be parallel to~$\Pi$). Otherwise,
the mirror symmetry of the charge distribution would imply that there
must be two distinct electric field vectors at~$P$ that have equal and
opposite components perpendicular to the plane, a contradiction.

Further, for any point~$P$ that lies on a line~$L$ that is the
intersection of two non-parallel mirror planes of the same charge
distribution, the electric field at~$P$ must be parallel to the
line~$L$.  This follows since the electric field vector at any such
point has to be parallel to two non-parallel planes at the same time,
which is possible only if the vector is parallel to their line of
intersection. For example, in
Fig.~\ref{fig:symmetry-planes-and-lines-of-cubic-surface}(c), the
mirror plane~$y=0$ that contains the green square intersects the
mirror plane~$x=0$ that contains the purple square at the line
corresponding to the $z$~axis, and so the electric field anywhere on
the $z$ axis is parallel to~$\hat{\bf z}$, and so has the form ${\bf
  E}(0,0,z) = (0,0,E_z)$. Similarly, in
Fig.~\ref{fig:symmetry-planes-and-lines-of-cubic-surface}(d), we see
that the mirror plane containing the blue rectangle~$ACGE$ and the
mirror plane containing the orange rectangle~$ADGF$ intersect at a
line that contains the diagonal line segment~$AOG$, and so the
electric field anywhere on the line passing through~$AOG$ is parallel
to this line.

If a charge distribution has three mutually non-parallel mirror planes
that intersect at some point~$P$, the electric field must be zero
at~$P$. This follows since the electric field at~$P$ has to be
parallel to three distinct mirror planes, which is only possible if
the vector is the zero vector.
For example, we see in
Fig.~\ref{fig:symmetry-planes-and-lines-of-cubic-surface}(c) that the
center of the cubic surface~$O$ lies at the intersection of the mirror
planes~$x=0$, $y=0$, and~$z=0$ and so the electric field at~$O$ lies
entirely within each of these planes simultaneously which requires all
three of its components to be zero.

A last observation is that, because an electric field varies
continuously except where the charge density changes discontinuously
(for the charged cubic surface, ${\bf E}$ is continuous along any path
that does not cross the surface), the internal electric field of the
charged cubic surface close to a mirror plane must be almost parallel
to that plane. Similarly, the electric field close to the intersection
of two mirror planes must be nearly parallel to their line of their
intersection, and the electric field near a point that is common to
three distinct mirror planes must be close to zero in magnitude. So
symmetry and continuity substantially constrain the qualitative
properties of the electric field near regions of symmetry associated
with the cubic surface.

\subsection{The interior electric field near the midpoints of faces
  points inwards, towards the cube's center}

\label{subsection-E-points-inwards-near-midpoints-of-faces}

Now that we understand that the electric field at a point on a mirror
plane is parallel to the mirror plane, we begin to obtain a
qualitative understanding of the electric field inside a uniformly
charged cubic surface. We claim that the internal electric field close
to any midpoint of a face must point inwards towards the center~$O$ of
the cubic surface. When this insight is combined with Gauss's law in
the next subsection, we will see that there must be places inside the
cubic surface where the electric field points outwards, from the
cube's center towards the cube's edges and corners.

To simplify some estimates that we make in the following discussion,
we assume that the cubic surface has unit length and unit surface
charge density in SI~units so that $L=1\,\rm m$, and~$\sigma = 1\,\rm
C/m^2$.  With these values, the total charge on any face of the cubic
surface is~$Q=\sigma L^2 = 1\,\rm C$ and the total charge of the cubic
surface is\cite{LargeChargeComment}~$Q_{\rm tot}= 6Q = 6\,\rm C$. We
will also occasionally not indicate the physical units when referring
to spatial coordinates, which should be understood to always be in
units of meters.

To understand why the internal electric field must point inwards near
the center of any face, we use our knowledge of the electric field of
a point charge and of the electric field of an infinite uniformly
charged plane\cite{Knight2012}. In
Fig.~\ref{fig:symmetry-planes-and-lines-of-cubic-surface}(c), let us
consider a point~$P$ that lies just above the bottom midpoint~$M_2 =
(0,0,-1/2)$ on the line segment~$M_1OM_2$. If~$P$ is sufficiently
close to~$M_2$, the electric field at~$P$ due to the bottom
face~$EFGH$ will approximately be equal to the electric field of an
infinite plane with uniform charge density~$\sigma$. (See for example
page~762 of Ref.~\onlinecite{Knight2012}, where this is shown
analytically to be the case for a point sufficiently close to and
above the center of a uniformly charged disk.) Since the charge
density~$\sigma$ is positive, we conclude that the electric
field~${\bf E}(P)_{EFGH}$ at~$P$ due to the face~$EFGH$ is given
approximately by
\begin{equation}
  \label{eq-E-EFGH}
  {\bf E}(P)_{EFGH} \approx {\sigma \over 2 \epsilon_0} \, \hat{\bf z} 
  \approx 2\pi K\, \hat{\bf z} ,
\end{equation}
where we used the fact that the vacuum permittivity~$\epsilon_0$ and
Coulomb's constant~$K$ are related by~$1/\epsilon_0 = 4 \pi K$. To one
significant digit in multiples of~$K$, the electric field at~$P$ due
to the bottom face has magnitude~$E \approx 6K$ and points in the
positive $z$-direction, away from the charged face and towards~$O$.

Now let us consider the contribution~${\bf E}(P)_{ABCD}$ to the
electric field at~$P$ from the top face~$ABCD$ in
Fig.~\ref{fig:symmetry-planes-and-lines-of-cubic-surface}(c). Since~$P$
lies on the intersection of two distinct mirror planes of the
face~$ABCD$, the electric field at~$P$ due to~$ABCD$ must be parallel
to the $z$ axis. Further, since~$ABCD$ is covered with a positive
charge density, we conclude that ${\bf E}(P)_{ABCD}$ must point in the
$-\hat{\bf z}$ direction. Now if square~$ABCD$ were extended to be an
infinite uniformly charged plane~$z=1/2$ with density~$\sigma$ that
contains square~$ABCD$, then the electric field at~$P$ due to this
plane would be exactly opposite in direction and equal in magnitude to
the electric field due to the bottom face and the electric field
at~$P$ would be approximately zero. But since~$ABCD$ is a finite
portion of an infinite charged plane, the electric field at~$P$ due to
$ABCD$ must be smaller in magnitude than the electric field
magnitude~$2\pi K$ of an infinite plane with the same charge
density. We conclude that the total electric field at a point~$P$ that
is sufficiently close to~$M_2$ and that is due to the top and bottom
faces of the cube must point upwards, towards the cube's center~$O$.

We can get a quick estimate of the approximate magnitude of ${\bf
  E}(P)_{ABCD}$ by approximating the distributed charge on the
face~$ABCD$ with a single point charge with total charge~$Q=1\,\rm C$
at the center~$M_1$ of this face. Coulomb's law then tells us that the
magnitude of the electric field at~$P$ due to~$ABCD$ is
\begin{equation}
  \label{eq-E-from-M1-point-charge}
  E(P)_{ABCD} 
   \approx {K Q \over L^2}
   = { K \left[L^2 \sigma\right]  \over L^2 }
   = K \sigma
   = K ,
\end{equation}
since we have assumed~$\sigma = 1\,\rm C/m^2$.  

This simple estimate is too big for two reasons. (And indeed, the
quantitative calculations of Section~\ref{sec:quantitative} show that
the value of the electric field at~$M_2$ due to the top face is~$0.8K$
to one digit, so that the estimate
Eq.~(\ref{eq-E-from-M1-point-charge}) is too large by about 20\%.)
First, we can think of approximating the face~$ABCD$ with a single
point charge~$Q=1\,\rm C$ at its center as being achieved by
relocating each infinitesimal point charge~$dq$ on that face in turn
to the center of that face. But changing the position of a point
charge~$dq$ to the face's center moves the charge closer to~$P$
because the center of that face is closer to~$P$ than all the other
points of the face. Since smaller distances~$d$ in Coulomb's law ${\bf
  E} \propto 1/d^2$ imply bigger electric fields, relocating all the
point charges on a face to its center makes the estimated total
electric field at point~$P$ larger than the actual value. Relocating a
point charge~$dq$ on~$ABCD$ to the face's center also changes the
orientation of the electric field vector at~$P$ due to~$dq$ to be more
parallel to the $z$-axis, which increases the component along the $z$
axis compared to its original component.

We now know that the total field at a point~$P$ near~$M_2$ due to only
the top and bottom faces gives an upwards vertical electric field of
approximate magnitude~$6K - 1K \approx 5K$.  However, since a
point~$P$ near~$M_2$ lies below the symmetry plane~$z=0$ that divides
the cubic surface horizontally in half, there are more infinitesimal
charges on each vertical side that lie above~$P$ than below~$P$. This
implies that the net electric field at~$P$ due to the four side faces
must have a net downwards $z$-component that also needs to be
considered when estimating the total electric field vector~${\bf
  E}(P)$ at~$P$.

\begin{figure}
\centering
\includegraphics[width=0.7\textwidth]{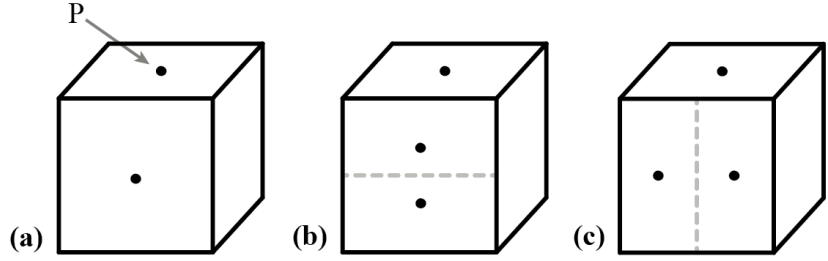}
\caption{The electric field~${\bf E}$ at a point~$P$ at the midpoint
  of a charged cubic surface (here the top) due to a side face (here
  the front face) can be estimated by replacing the distributed charge
  of the front face with one or more point charges. In (a), a single
  point charge with~$Q=1\,\rm C$ at the center of the face is
  used. Panels~(b) and~(c) show two ways that the charge on the front
  panel could be approximated by two equal point charges with
  charge~$Q=1/2\,\rm C$, by dividing the face into two equal smaller
  rectangles and replacing the charges on each rectangle by a point
  charge at its midpoint. }
\label{fig:approximating-side-faces-with-point-charges}
\end{figure}

We can again obtain a quick estimate by replacing each vertical side
face with a single point particle of charge~$Q=1\,\rm C$ at its center
as shown in
Fig.~\ref{fig:approximating-side-faces-with-point-charges}(a) and by
adding up the four electric field vectors at~$P$ due to these four
point charges.  It is harder now to determine whether this
single-point-charge approximation will cause an overestimate or
underestimate of the exact value~${\bf E}(P)_{\rm side-faces}$ since,
upon relocating an infinitesimal charge~$dq$ on a side face to the
center of that side face, the distance to~$P$ can become larger or
smaller, and the electric field vector at~$P$ due to~$dq$ can become
more or less parallel to the $z$ axis. 

In any case, by approximating the four side faces with point
charges~$Q=1\,\rm C$ at their centers, by defining~${\bf X}_P =
(0,0,-1/2)$ to be the position vector of point~$P$ at the bottom
midface and ${\bf X}_{F_1} = (1/2,0,0)$ to be the position vector of
the point charge at the front midface of the cubic surface, by
defining~$d_{PF_1}=\|{\bf X}_P - {\bf X}_{F_1}\| = 1/\sqrt{2}$ to be
the distance between the two vectors, by defining $\hat{\bf r}_{PF_1}
= ({\bf X}_P - {\bf X}_{F_1})/d_{PF_1} = (-1/\sqrt{2},0,-1/\sqrt{2})$
to be the unit vector that points from the front midface to~$P$, and
finally by observing that, by symmetry, the final $z$-component of the
electric field at~$P$ due to the four sides is four times the $z$
component from any one side, we find that
\begin{equation}
  \label{eq-E-from-four-point-charges-on-side-faces}
  {\bf E}(P)_{\rm side-faces}
  \approx  4 \left( 
     \hat{\bf z} \cdot { K Q \over d_{PF_1}^2 } \hat{\bf r}_{PF_1} \right)
  \approx - 4 \sqrt{2} K \, \hat{\bf z}
  \approx - 5.7 K \, \hat{\bf z} . 
\end{equation}
So by approximating the bottom face with an infinite plane and the
other five side faces with point charges~$Q=1 \,\rm C$ at their
centers, we get a total electric vector near~$M_2$ in the~$-z$
direction of length~$2\pi K - 1.0K - 5.7K \approx -0.4K$ pointing
downwards. This suggests that the electric field just inside the cubic
surface and near the midpoint of a face points outwards, away from the
origin~$O$.

However, since the estimate~$E_z \approx (2\pi - 1.0 + 5.7)K \approx
-0.4K$ involves cancellations of estimates bigger than~$K$ in
magnitude to give a final answer smaller than~$K$ in magnitude, we
need to worry about whether using a point charge to approximate the
electric field of a charged face is sufficiently accurate. Since the
four side faces contribute the second largest amount to the total
electric field at~$P$, it would be useful to explore whether a more
careful approximation of the electric field from each side face might
affect the sign of~$E_z$. A next step that is easy to compute would be
to approximate the surface charge of each side face with two equal
point charges with values~$q=Q/2=1/2\,\rm C$ as shown in panels~(b)
and~(c) of
Fig.~\ref{fig:approximating-side-faces-with-point-charges}. For panel
Fig.~\ref{fig:approximating-side-faces-with-point-charges}(b), the
total electric field at~$P$ due to the eight point charges of the
sides is given by
\begin{equation}
  \label{eq-estimate-two-charges-per-side-face}
  E_z(P)_{\rm 8-point-charges} 
   \approx  4 \, \hat{\bf z} \cdot \left(
    { K (Q/2) \over d_{PF_1}^2 } \hat{\bf r}_{PF_1} +
    { K (Q/2) \over d_{PF_2}^2 } \hat{\bf r}_{PF_2}
  \right)
  \approx - 4.9 K ,
\end{equation}
to two digits. Here we define~${\bf X}_{F_1}=(1/2,0,1/4)$ and~${\bf
  X}_{F_2} = (1/2,0,-1/4)$ to be the position vectors of the two point
charges with charge~$Q/2 = (1/2)\,\rm C$ on the front face in
Fig.~\ref{fig:approximating-side-faces-with-point-charges}(b), and
then, as before, we define~$d_{PF_1} = \|{\bf X}_P - {\bf X}_{F_1}\|$,
$\hat{\bf r}_{PF_1} = ({\bf X}_P - {\bf X}_{F_1})/d_{PF_1}$, and
similarly for $d_{PF_2}$ and~$\hat{\bf r}_{PF_2}$.  A calculation
similar to Eq.~(\ref{eq-estimate-two-charges-per-side-face}) for
Fig.~\ref{fig:approximating-side-faces-with-point-charges}(c) gives a
value of $E_z \approx -4.7K$ so both arrangements of two equal charges
per side lead to the same conclusion, that the magnitude of~${\bf
  E}(P)$ due to all six faces is about $(2\pi - 1.0 - 4.9)K \approx
0.4K$ pointing in the $+\hat{\bf z}$ direction, which represents a
reversal in direction of the electric field near~$M_2$ compared to a
one-charge-per-face approximation.

One would presume that this new estimate using two charges per face is
more accurate than an estimate based on one charge per face since two
points charges per face should do a better job of getting the
magnitude and direction of the face's electric field at~$M_2$ correct.
Other calculations using more point charges per face indeed confirm
that the electric field near the midpoint of a face indeed points
towards the center. It is possible to calculate the exact total
electric field at a point~$P$ quite close to~$M_2$ (see
Appendix~\ref{appendix:exact-electric-field}) and one finds that~${\bf
  E}(P) \approx 1.9K \hat{\bf z}$ to two digits. The exact
contribution to the electric field at~$M_2$ due to the side faces is
therefore $3.6K$ in the $-\hat{\bf z}$ direction, compared to the
estimates of~$4.9K$ or~$4.7K$ for panels~(b) and~(c) of
Fig.~\ref{fig:approximating-side-faces-with-point-charges}. So two
point charges per side face get the sign right but produce an error in
magnitude of the electric field at~$P$ of about 40\%.

We conclude that the internal electric field points towards the center
of the cubic surface for points sufficiently close to the midpoint of
any face. From this, we can deduce what is the qualitative form of the
electric field~${\bf E}$ along the entire line segment~$M_1OM_2$. (By
symmetry, this will also be the qualitative form of the electric field
along the other two line segments connecting midpoints of opposite
faces.) The symmetry arguments of
Sec~\ref{subsection-E-parallel-to-mirror-planes} imply that~${\bf E}$
at any point on~$M_1OM_2$ is parallel to~$M_1OM_2$ and so~${\bf E}$
must have the form~$(0,0,E_z)$ on this line segment.  The
$z$-component~$E_z(z)$ is positive near~$M_2$ and, by symmetry, must
be negative near~$M_1$ and we further know that~$E_z(0)=0$ since the
center of the cube lies at the intersection of at least three distinct
mirror planes and so~${\bf E}$ must vanish there. We thus expect
$E_z(z)$ to be a smoothly varying odd function of~$z$, $E_z(-z) = -
E_z(z)$, that is positive for~$z < 0$ and that decreases through zero
to negative values for~$z > 0$. Further, the magnitude of~$E_z$
along~$M_1OM_2$ must always be less than the magnitude~$2\pi K \approx
6K$ of the electric field produced by an infinite plane with surface
charge density~$\sigma =1\,\rm C/m^2$. The quantitative calculation
Fig.~\ref{fig:E-V-along-symmetry-lines}(a) of
Section~\ref{sec:quantitative} over the range $-1/2 \le z \le 1/2$
shows that this qualitative thinking is correct.

\subsection{Gauss's law implies that there are places where the
  interior electric field points outwards, towards edges and towards
  vertices}
\label{subsection-application-of-gauss-law}

\begin{figure}
\centering
\includegraphics[width=0.3\textwidth]{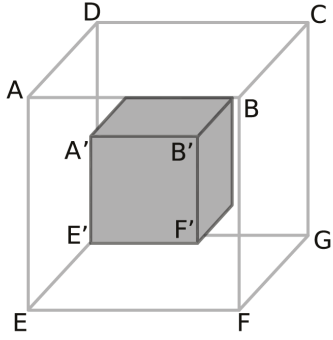}
\caption{ (a) The cubic surface $S=A'B'C'D'E'F'G'H'$ is a Gaussian
  surface that is concentric with the charged cubic surface $ABCDEFGH$
  and that lies just within the charged cubic surface. The total
  charge enclosed by~$S$ is zero which implies, by the symmetry of the
  cubic surface, that the flux through each face such as $A'B'F'E'$ is
  zero. }
\label{fig:cubic-Gaussian-surface}
\end{figure}

We now combine the insight that the internal electric field points
inwards near the center of faces with a qualitative application of
Gauss's law to deduce that there have to be locations inside the
charged cubic surface where the electric field points away from the
cube's center~$O$. Consider a Gaussian cubic
surface~$S=A'B'C'D'E'F'G'H'$ that is concentric with and lies just
inside the charged cubic surface $ABCDEFGH$ as shown in
Fig.~\ref{fig:cubic-Gaussian-surface}. Because there is no charge
inside the surface~$S$, Gauss's law gives:
\begin{equation}
  \label{eq-Gauss-law-applied-to-cubic-surface}
  { Q_{\rm enclosed} \over \epsilon_0 }
   = 0 
   = \Phi_{\rm total} 
   = \int_S {\bf E} \cdot d{\bf A} 
   = 6 \int_\Box {\bf E} \cdot d{\bf A}
   = 6 \Phi_\Box ,
\end{equation}
so the flux~$\Phi_\Box$ through any face~$\Box$ of the cubic
surface~$S$ is zero. Here we have used the symmetry of the cube to
deduce that the flux~$\Phi_\Box$ through any face must be the same so
the flux integral~$\Phi_{\rm total}$ over the surface~$S$ is six times
the flux~$\Phi_\Box$ through any one face.

If we consider a particular face of the surface~$S$, say the front
face~$A'B'F'E'$ that lies at a coordinate~$x_0$ that is just less
than~$x=(1/2)\,\rm m$, the flux integral for that face becomes a
two-dimensional integral of the $x$-component~$E_x(x_0,y,z)$ of the
electric field vector since the face~$A'B'E'F'$ is perpendicular to
the $x$~axis:
\begin{equation}
  \label{eq-flux-integral-front-face}
  0 = \Phi_\Box
    = \Phi_{A'B'F'E'}
    = \int_{A'B'F'E'} {\bf E} \cdot d{\bf A}
    = \int_{A'B'F'E'} {\bf E} \cdot \left( dA \, \hat{\bf x} \right)
    = \int_{A'B'F'E'} E_x \, dy\,dz .
\end{equation}
The last integral $\int E_x \, dx \, dy$ can be thought of as the
limit of a finite sum of values $E_x(x_0,y_i,z_i) \Delta{A}_i$ over
some fine uniform grid of tiny identical square areas~$\Delta A_i$
that all have the same area $dA_i = \Delta{y}\,\Delta{z} =
\Delta{A}$. So the flux integral
Eq.~(\ref{eq-flux-integral-front-face}) can be thought of as
approximately equal to~$\Delta{A} \sum_i E_x(x_0,y_i,z_i)$, i.e., it
is proportional to the sum of the $x$-components of the electric field
values over the face~$A'B'F'E'$. Since this sum is zero by
Eq.~(\ref{eq-flux-integral-front-face}), we conclude that the values
of~$E_x$ cannot be everywhere positive or everywhere negative, else
the sum $\sum_i E_x(x_0,y_i,z_i)$ would be respectively positive and
negative, a contradiction.

But we know from
Sec.~\ref{subsection-E-points-inwards-near-midpoints-of-faces} that,
near the center of the face~$ABFE$ of the charged cubic surface, the
interior electric field points inwards towards the origin~$O$. Since
the face~$ABFE$ is perpendicular to the $x$-axis and lies at the
coordinate~$x=1/2\,\rm m$, this specifically implies that $E_x$ must
be negative near the center of~$ABFE$. But the electric field~${\bf
  E}$ varies continuously everywhere inside the cubic surface. (Only
along a path that crosses the charged surface would ${\bf E}$ change
discontinuously.) Provided that the face~$A'B'F'E'$ is sufficiently
close to~$ABFE$, continuity of~$E_x$ implies that~$E_x$ must also be
negative over some finite region near the center of the
face~$A'B'F'E'$.  But then the only way that
Eq.~(\ref{eq-flux-integral-front-face}) can hold is for~$E_x$ to be
positive on $A'B'F'E'$ in regions away from the middle of the face so
that the negative and positive values of~$E_x$ over the entire face
add to zero.  We conclude that there must be points inside the charged
cubic surface where the electric field points away from the center of
the cube. This immediately implies that the interior electric field
must have a complicated structure, pointing inwards in some locations
(near the middle of each face) and outwards in other locations.

A simple way that~$E_x$ could be negative near the middle
of~$A'B'F'E'$ and positive away from the middle consistent with the
symmetry of a cube would be for~$E_x$ to be negative in some
square-like region near the face center and positive elsewhere on the
face. The quantitative calculations of Sec.~\ref{sec:quantitative}
show that this simplest case is what actually occurs, see
Fig.~\ref{fig:flux-through-front-face} below.

If we assume this simplest case, then we can understand qualitatively
that the electric field inside the cubic surface must actually point
towards edges or towards corners for points near edges or near
corners. For example, for interior points close to the points~$A'$,
$B'$, $F'$, and~$E'$ in Fig.~\ref{fig:cubic-Gaussian-surface}, the
components of~${\bf E}$ point outwards on three planes parallel to
respective~$y=0$, $x=0$, and~$z=0$. This tells us that the electric
vectors near these points must point outwards towards the
corners. These qualitative insights are confirmed by our quantitative
discussion in Section~\ref{sec:quantitative}.

The qualitative arguments in this section only hold for interior
Gaussian cubic surfaces~$A'B'C'D'E'F'G'H'$ whose faces are
sufficiently close to the faces of the charged cubic surface. How the
electric field varies more deeply in the cube's interior cannot be
worked out qualitatively and one has to turn to quantitative
calculations to understand the bigger picture. The quantitative
calculations show that the arguments of this subsection hold generally
for all interior cubic Gaussian surfaces: everywhere in the interior,
the electric field points inwards along the faces of a Gaussian cube
and outwards near the edges and corners of the Gaussian cube.

\subsection{A qualitative comparison of the non-conducting cubic surface with three
similar problems}
\label{subsect-related-systems-with-zero-field}

Before discussing our quantitative results, we compare our qualitative
conclusions about the nonzero electric field inside a uniformly
charged non-conducting cubic surface with three related problems for
which the electric field inside some interior region of a symmetric
charge distribution is zero.  This comparison helps to clarify why the
electric field inside the non-conducting charged cubic surface is
nonzero.

First we ask: why is the electric field nonzero inside the charge
distribution~$\sigma_{\rm cube}$ consisting of a uniformly charged
cubic surface while the electric field is zero everywhere inside the
charge distribution~$\sigma_{\rm sphere}$ consisting of a uniformly
charged spherical surface?  At first glance, these seem to be similar
problems since the charge distributions are both highly symmetric.

One answer is that a spherical surface has many more symmetries than a
cubic surface. For example, instead of having nine mirror planes
like~$\sigma_{\rm cube}$ (see
Fig.~\ref{fig:symmetry-planes-and-lines-of-cubic-surface}), the
distribution~$\sigma_{\rm sphere}$ has infinitely many mirror planes
since any plane passing through the center of the spherical surface is
a mirror plane of~$\sigma_{\rm sphere}$. These infinitely many mirror
planes allow one to show\cite{EisFnOfRadius} that the electric
field~${\bf E}$ inside~$\sigma_{\rm sphere}$ is radial, ${\bf E}=E
\hat{\bf r}$, and further that the electric field magnitude~$E$
depends only on radius,~$E=E(r)$. These two facts then lead to the
usual argument given in introductory physics
textbooks\cite{Knight2012}, that the flux integral~$\Phi = \int {\bf
  E} \cdot d{\bf A}$ in Gauss's law, applied to a spherical Gaussian
surface of radius~$r$ concentric with and inside the uniformly charged
spherical surface, becomes a simple product~$\Phi = E(r) A$ and so the
vanishing of the flux (since there is no charge inside the spherical
Gaussian surface) implies the vanishing of the electric field. In
contrast, the charge distribution~$\sigma_{\rm cube}$ does not have
enough symmetry for one to conclude that the interior electric field
is radial (which it isn't), so the flux integral cannot be written as
the product of some area times some constant electric field magnitude.

\begin{figure}
\centering
\includegraphics[width=0.3\textwidth]{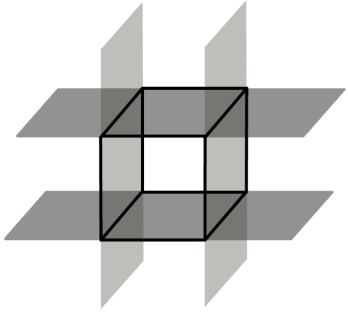}
\caption{A charge distribution consisting of three identical pairs of
  parallel planes, each with a uniform surface charge
  density~$\sigma$. The front and rear pair of planes are not shown to
  make the diagram easier to understand. The electric field is zero
  everywhere inside the central cubic region since the electric fields
  of opposing planes cancel exactly at any interior point.}
\label{fig:cube-formed-by-three-pairs-charged-planes}
\end{figure}

We next consider an empty cubic region that lies within a symmetric
charge distribution that consists of three identical parallel pairs of
uniformly charged infinite non-conducting planes, see
Fig.~\ref{fig:cube-formed-by-three-pairs-charged-planes}. This figure
can be obtained by extending each face of the cubic surface in
Fig.~\ref{fig:charged-cubic-surface} to an infinite plane that
contains that face and that has the same surface charge
density~$\sigma$.

Since the electric field due to an infinite charged plane is uniform
on a given side of the plane and points away from the plane if~$\sigma
> 0$, we conclude that the electric field~${\bf E}$ must be zero
everywhere in the cubic region of
Fig.~\ref{fig:cube-formed-by-three-pairs-charged-planes} since, at any
point~$P$ inside the cube, the electric field vectors from opposing
planes are equal and opposite and so cancel exactly. The nonzero
internal electric field of Fig.~\ref{fig:charged-cubic-surface}
therefore arises from the finite size of the faces of the cubic
surface, which in turn implies that each face produces a nonuniform
electric field, that decreases in magnitude with increasing distance
from a face. Since a planar charge distribution in introductory
physics textbooks is often represented visually by a finite rectangle
contained in the plane, it is easy for students to conclude
incorrectly that the electric field inside the cubic surface
Fig.~\ref{fig:charged-cubic-surface} must be the same as that of
Fig.~\ref{fig:cube-formed-by-three-pairs-charged-planes}.

The third example that we consider is the electric field inside a
charged {\em conducting} cubic surface that has the same size and same
total charge as the charged non-conducting surface of
Fig.~\ref{fig:charged-cubic-surface}.  Introductory physics textbooks
discuss the fact that the electric field everywhere inside a hollow
conductor must be zero if that conductor, charged or not, is in
electrostatic
equilibrium\cite{Tipler2007,Serway2011,Young2011,Knight2012,Cutnell2014}.
So we have another situation that is confusing to student who are
learning about electrostatics: how can one have two identical cubic
surfaces with identical total charges, and yet one surface has a
nonzero interior electric field while the other surface has a zero
interior electric field?

The key insight here is that, on a conducting cubic surface, the
charges are mobile and move about (because of mutual repulsion) until,
in electrostatic equilibrium, the surface charge density~$\sigma$~is
nonuniform in just such a way that the surface and interior are all
equipotential\cite{Knight2012,Griffiths2012}. If there are no other
charges inside the surface, this implies a zero interior electric
field and that the external electric field is everywhere locally
perpendicular to the cubic surface (except at edges and
corners). Introductory physics textbooks mention briefly and
qualitatively\cite{Tipler2007,Knight2012,Young2011} that charged
non-spherical three-dimensional conductors in equilibrium have
nonuniform surface charge densities, and that these densities are
larger where the radius of curvature of the surface is smaller. But
these books and even the commonly used upper-level books on
electricity and magnetism\cite{Griffiths2012,Purcell2013} do not
discuss quantitative examples of nonuniform charge densities.


There is a complimentary insight, which is that the surface of a
uniformly charged non-conductor of non-spherical shape cannot be
equipotential. This follows since a charged non-spherical conductor
has a non-uniform surface charge density.

\begin{figure}
\centering
\includegraphics[width=0.7\textwidth]{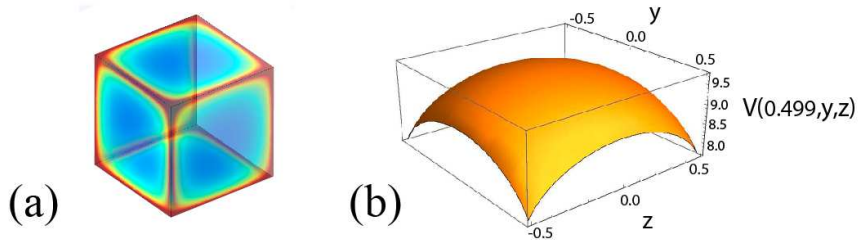}
\caption{(a) Color density plot of the nonuniform surface charge
  density~$\sigma$ on the faces of a conducting equipotential cubic
  surface with potential~$V= 1\,\rm V$, as calculated using a
  commercial computer code\cite{ComsolCalculation}. The density is
  approximately constant in the blue regions with value $\sigma
  \approx 8 \times 10^{-11}\,\rm C/m^2$ and increases near the edges
  (red brown regions) by a factor of about five.  (b) The
  non-conducting uniformly charged cubic surface is not equipotential,
  as shown by the surface plot of the potential~$V(0.499,y,z)$ of
  Eq.~(\ref{eq-V-cubic-surface-exact}) on the $yz$-face of a uniformly
  charged non-conducting cubic surface.  }
\label{fig:sigma-versus-V-nonconducting-vs-conducting-surfaces}
\end{figure}

Figure~\ref{fig:sigma-versus-V-nonconducting-vs-conducting-surfaces}
clarifies these two insights in the context of a cubic surface.
Fig.~\ref{fig:sigma-versus-V-nonconducting-vs-conducting-surfaces}(a)
shows a numerical approximation to the nonuniform surface charge
density~$\sigma$ for a charged conducting equipotential cubic surface
in electrostatic equilibrium\cite{ComsolCalculation}. (Note that
this~$\sigma$ is what the uniform charge density of
Fig.~\ref{fig:charged-cubic-surface} would evolve into if the
uniformly charged non-conducting cubic surface were to become
conducting.) For this calculation with a spatial resolution of
$120\times 120$ grid points per face, about 10\% of the area of each
face contains about 70\% of the total charge per face so most of the
surface charge ends up near the edges. As the spatial resolution
becomes finer, the surface charge density on the edges increases,
corresponding to the fact that the surface charge density
mathematically diverges to infinity where sharp edges occur on a
charged conductor\cite{Jackson1998}.

Fig.~\ref{fig:sigma-versus-V-nonconducting-vs-conducting-surfaces}(b)
shows the complimentary result of how the potential~$V$ is nonuniform
over one face of the uniformly charged non-conducting cube.  (The
potential~$V$ is easily evaluated numerically using the analytical
expressions Eqs.~(\ref{eq-V-analytical-expression-for-rectangle})
and~(\ref{eq-V-cubic-surface-exact}) of
Appendix~\ref{appendix:exact-electric-field}.) The potential~$V$
varies modestly in magnitude by about 30~percent, from about~$7.7\,\rm
V$ at the corners to about $9.8\,\rm V$ at the face's center.
However, it is the gradient of~$V$ that determines the electric field
via ${\bf E} = -\nabla{V}$ and it is not apparent in
Fig.~\ref{fig:sigma-versus-V-nonconducting-vs-conducting-surfaces}(b)
that the local slope (magnitude of the gradient) is becoming vertical
and so diverges at the corners of the non-conducting cubic
surface. This divergence is not easy to understand qualitatively, and
we demonstrate this fact with a short Mathematica\cite{Mathematica15}
calculation in Appendix~\ref{appendix:exact-electric-field}.

We note that the calculation of the surface charge density on a
charged cubic conductor is a difficult calculation that is not
discussed even in graduate textbooks on electricity and magnetism like
Jackson\cite{Jackson1998}. The authors only know of numerical
calculations using specialized computer codes that have been carried
out mainly by electrical engineers who have been interested in the
capacitance of a cube-shaped
capacitor\cite{Reitan1951,Hwang2004,Velickovic2004}. Nevertheless, the
fact that the surface charge density on a conducting cubic surface in
electrostatic equilibrium is not uniform is easily understood by
undergraduates and they should be familiar with at least one
quantitative example like the one discussed in this paper.

\section{Quantitative Insights}
\label{sec:quantitative}

In this section, we use an analytical expression for the electric
field~${\bf E}(x,y,z)$ of a uniformly charged cubic surface (see
Appendix~\ref{appendix:exact-electric-field}) to confirm and to extend
the qualitative results of the previous section. The expression
for~${\bf E}$ is so long (several pages) that it provides little
insight by itself. So a key contribution of this section is a
discussion of how to visualize the three-dimensional internal electric
field~${\bf E}$ by plotting it on line segments and on surfaces that
lie within mirror planes (see
Sect.~\ref{subsection-E-parallel-to-mirror-planes}). We can then use
the fact that the electric field varies continuously in space along
paths that do not cross a charged surface to extend knowledge from the
symmetry plots to other spatial regions.  We also discuss briefly the
properties of the electric potential~$V$ as revealed by several line
and surface plots.

We note that, for obtaining the results of this section, it was
convenient but not essential that an explicit mathematical expression
for~${\bf E}(x,y,z)$ was available. In fact, for general charge
distributions, even in one spatial dimension, one would not expect to
be able to obtain explicit mathematical expressions for the potential
or for the electric field since the integrals analogous to
Eq.~(\ref{eq-V-integral-for-rectangle}) in
Appendix~\ref{appendix:exact-electric-field} can generally not be
evaluated in closed form in terms of elementary functions.

Instead, we could have easily obtained all the results of this section
by using an elementary numerical method based on discretization and on
superposition, in which an arbitrary charge distribution~$\rho$ is
approximated by some finite set of point
charges\cite{NumMethodFirst}. (See
Sect.~\ref{subsection-E-points-inwards-near-midpoints-of-faces} where
we approximated some faces of the cubic surface with one or two point
charges, and also see the brief description of the general numerical
method in Appendix~\ref{appendix:exact-electric-field}.) Such a
numerical method has the advantage over analytical expressions of
simplicity and generality since all the details can be understood at
the freshman-physics level and the method applies to arbitrarily
complicated charge distributions. But one drawback of a numerical
method is that a student would have to know how to write a computer
program (or use a previously written program) to carry out a summation
over the point charges, and this can be challenging for charge
distributions of irregular shape such as the cubic surface. A second
drawback is that, for each calculation that uses a numerical method,
one has to explore how the accuracy of the result depends on the
number and locations of the point charges used to approximate the
charge distribution. In this paper, we avoided this last step by using
the exact expression for~${\bf E}$.

\begin{figure}
\centering
\includegraphics[width=0.4\textwidth]{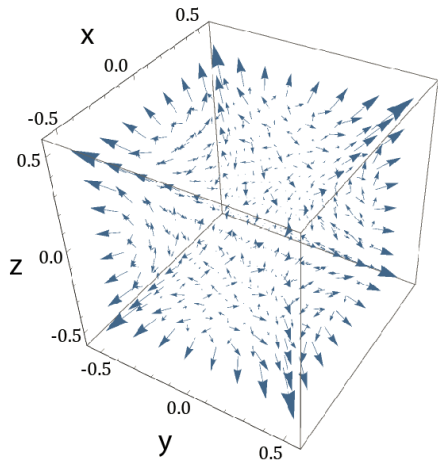}
\caption{Three-dimensional vector plot of the electric field~${\bf E}$
  inside the uniformly charged non-conducting cubic surface, over the
  region $|x|, |y|, |z| < 0.45$. (The plot was created using the
  Mathematica command {\tt VectorPlot3D}.) The geometry of~${\bf E}$
  is difficult to determine from this plot since interior vectors are
  obscured by vectors closer to the viewer.}
\label{fig:3d-vector-plot-inside-cubic-surface}
\end{figure}

Whether an exact expression or a numerical method that approximates
the exact field is used, there is still the challenge of how to
visualize and to understand three-dimensional spatially varying vector
fields. One might think that a first step would be to create a vector
plot of~${\bf E}$ on some regular 3D~grid of points like
Fig.~\ref{fig:3d-vector-plot-inside-cubic-surface}, but this is not
helpful since the vectors near the front of the plot partially hide
the vectors that are further back, and the orientations of small
vectors are difficult to determine visually. Instead, we have found it
useful to plot~${\bf E}$ on line segments and on rectangular regions
that lie within mirror planes of the cubic surface since, as discussed
in Sect.~\ref{subsection-E-parallel-to-mirror-planes}, the electric
field is parallel to such regions so such one- and two-dimensional
plots provide complete knowledge of the electric field.

\subsection{Confirmation of the qualitative insights of
  Sect.~\ref{sec:qualitative}}
\label{subsec-confirmation-qualitative-insights}

\begin{figure}
\centering
\includegraphics[width=0.7\textwidth]{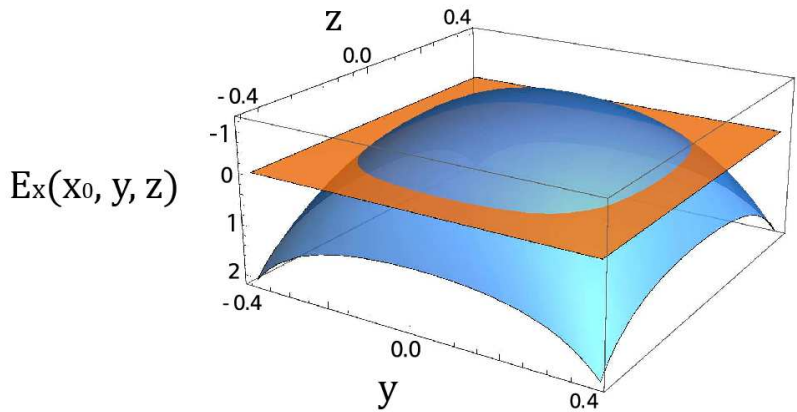}
\caption{Surface plot of the $x$-component~$E_x(x_0,y,z)$ of the
  electric field vector on the front face~$x_0=0.4L$ of a Gaussian
  cubic surface centered on the origin. This plot is proportional to
  the local flux~$d\Phi = E_x \Delta{A}$. The orange plane indicates
  where~$E_x$ has the value zero, so the surface above the plane in
  the middle is where the flux is negative (the electric field points
  into the Gaussian surface), and the surface below the plane is where
  the flux is positive (electric field points out of the Gaussian
  surface). The total flux through this front face is zero.  }
\label{fig:flux-through-front-face}
\end{figure}

We begin our quantitative discussion by using the exact expression
for~${\bf E}(x,y,z)$ in Appendix~\ref{appendix:exact-electric-field}
to show in Fig.~\ref{fig:flux-through-front-face} how the local
electric flux~$d\Phi(x_0,y,z) = {\bf E}(x_0,y,z)\cdot d{\bf
  A}(x_0,y,z) = E_x(x_0,y,z) \, \Delta{y} \, \Delta{z}$ varies over
the front face~$A'B'F'E'$ of an interior cubic Gaussian surface (see
Fig.~\ref{fig:cubic-Gaussian-surface}) whose sides have length
$0.8\,\rm m$. Figure~\ref{fig:flux-through-front-face} shows a surface
plot of~$E_x(x_0,y,z)$ over the region $|y|, |z| \le 0.4\,\rm m$. The
middle bulge above the orange plane where~$E_x=0$ denotes where the
electric field points inwards (where~$E_x$ is negative).

This figure confirms the qualitative conclusions of
Sects.~\ref{subsection-E-points-inwards-near-midpoints-of-faces} and
of~\ref{subsection-application-of-gauss-law} that the interior
electric field near the midpoints of the faces points inwards.  But
now we understand quantitatively how the flux is zero over this face:
the component~$E_x$ is negative in a larger central area but with
smaller magnitude~$|E_x|$, while the component~$E_x$ is positive in a
smaller area but with a larger magnitude outside the central
region. The larger number of smaller negative values balance the
smaller number of larger positive values, giving a net flux of zero.

\begin{figure}
\centering
\includegraphics[width=1.0\textwidth]{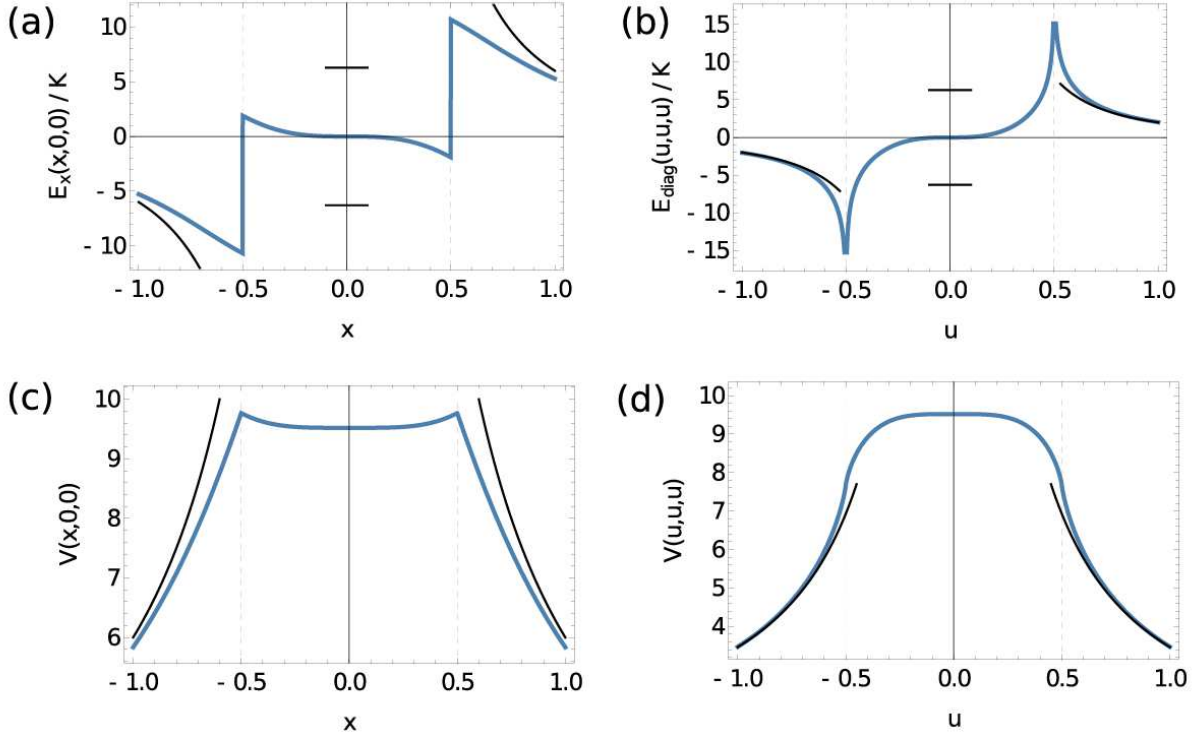}
\caption{ (a) Plot of $E_x(x,0,0)/K$ versus~$x$ along the $x$-axis
  where~$K$ is Coulomb's constant. The two thin black curves are the
  electric field component $E_x/K = \pm 6/x^2$ at location~$(x,0,0)$
  for a point charge~$Q=6\,\rm C$ at the origin. The two horizontal
  tick marks at $E/K = \pm 2 \pi \approx \pm 6.3$ on the~$x=0$
  vertical line denote the magnitude of the electric field for an
  infinite plane with the same charge density~$\sigma =1$. (b) Plot of
  the electric field component~$E_{\rm diag} / K = [{\bf E}(u,u,u)
    \cdot \hat{\bf n}]/K$ parallel to the cube diagonal~${\bf x}(s) =
  s \hat{\bf n}$ where~$\hat{\bf n} = (1,1,1)/\sqrt{3}$. (c)
  Potential~$V(x,0,0)$ plotted along the $x$-axis. The thin black
  lines correspond to the potential~$V = 6/x$ associated with a point
  charge~$Q=6\,\rm C$ placed at the center of the cubic surface. (d)
  Potential~$V(u,u,u)$ plotted along the same diagonal as in
  panel~(b).  }
\label{fig:E-V-along-symmetry-lines}
\end{figure}

Panels~(a) and~(b) of Figure~\ref{fig:E-V-along-symmetry-lines} next
show how the electric field varies quantitatively along two symmetry
lines of the charged cubic
surface. Fig.~\ref{fig:E-V-along-symmetry-lines}(a) shows how~${\bf E}
= E_x \hat{\bf x}$ varies along the line that passes through the two
opposing midpoints~$(x,y,z)=(1/2,0,0)$ and~$(-1/2,0,0)$. This plot
confirms the earlier qualitative conclusion that the electric field
points inwards near the midpoints of faces and vanishes at the
center~$O$, and shows further that the electric field has a small
magnitude over much of the interior of cubic surface. (Note the flat
approximately zero behavior of~$E_x$ for $|x| \lesssim 0.3$.) The
magnitude of the interior electric field on this symmetry line is
everywhere smaller than the electric field field magnitude~$2\pi K
\sigma \approx 6.2K$ of an infinite plane with the same charge density
(the two horizontal tick marks on the $x=0$ vertical axis denote this
magnitude). As one proceeds along this symmetry line from just inside
to just outside the cubic surface (so $|x|$ increases from just less
than~1/2 to just greater than~1/2), the electric field magnitude~$E$
changes discontinuously to a finite value that is larger than the
electric field magnitude of an infinite plane of the same charge
density. That the electric field magnitude is larger just outside the
surface was explained in
Sect.~\ref{subsection-only-interior-field-interesting}.

Figure~\ref{fig:E-V-along-symmetry-lines}(b) shows a similar plot
except along a symmetry line ${\bf x}(u) = (u/2,u/2,u/2)$ that
connects the diagonally opposite corners $(-1/2,-1/2,-1/2)$
and~$(1/2,1/2,1/2)$. From the sign of the electric field
component~$E_{\rm diag} = {\bf E} \cdot \hat{\bf n}$ along this line,
we see that, inside the cubic surface, the electric field everywhere
except at the origin points {\em outwards} towards the corners, in
agreement with our qualitative discussion in
Sect.~\ref{subsection-application-of-gauss-law}. We further see that
the electric field magnitude diverges to infinity at the corners,
which we explain briefly mathematically in
Sect~\ref{subect-log-divergence} of the appendix. In contrast, the
electric field is finite in magnitude at the midpoints of the cubic
surface (Fig.~\ref{fig:E-V-along-symmetry-lines}(a)).

In panels~(a) and~(b) of Fig.~\ref{fig:E-V-along-symmetry-lines}, we
also show by thin black curves the electric field corresponding to a
point particle at~$O$ with the same total charge~$Q=6\,\rm C$ as the
cubic surface. We expect the external electric field of the cubic
surface to converge to that of the point charge for distances
sufficiently far from the origin, but the quantitative calculation
shows the surprising result that the external field already acts
accurately like that of a point charge for distances that are as close
as one cube side~$L$ from the center.

\begin{figure}
\centering
\includegraphics[width=0.8\textwidth]{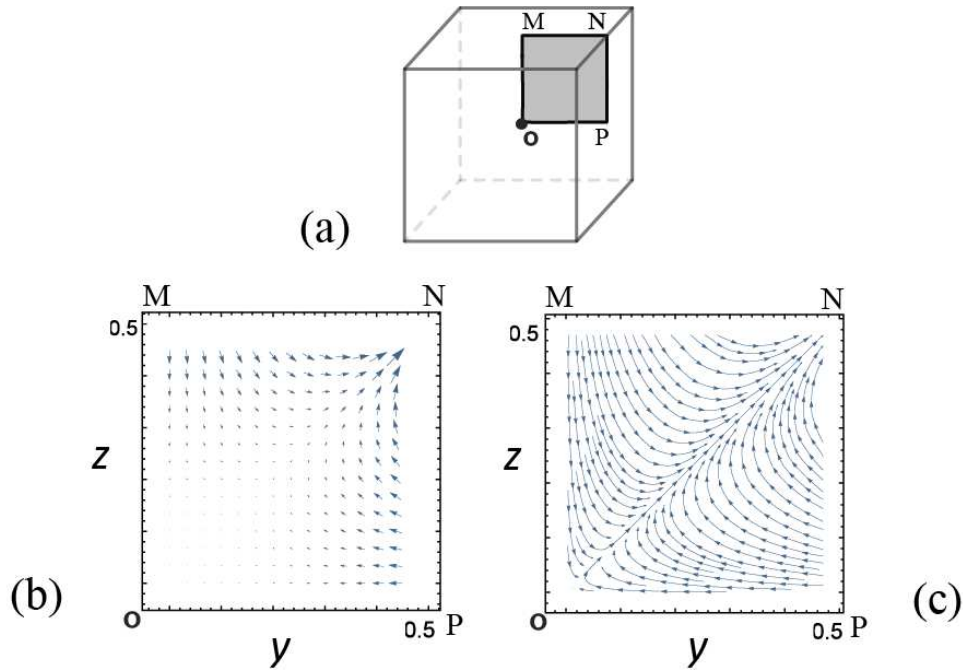}
\caption{(a) Schematic showing the relation of the plotting region,
  square~$MNOP$ in the mirror plane~$x=0$, to the charged cubic
  surface. (b) Vector plot of the electric field~${\bf E}(0,y,z) =
  (0,E_y,E_z)$ over the square region~$MNOP$ defined by~$x=0$ and by
  $0 \le y < 1/2$, $0 \le z < 1/2$. The electric field points inwards
  near the centers~$M$ and~$P$ of the top and side faces, and point
  outwards towards the edge that passes perpendicular to the
  point~$N$. (c) A streamline plot of~${\bf E}$ over the same region
  reveals the geometry of~${\bf E}$ everywhere in the interior. This
  plot was created using the Mathematica command {\tt StreamPlot}.}
\label{fig:vector-streamline-plot-x=0-plane}
\end{figure}

\begin{figure}
\centering
\includegraphics[width=.8\textwidth]{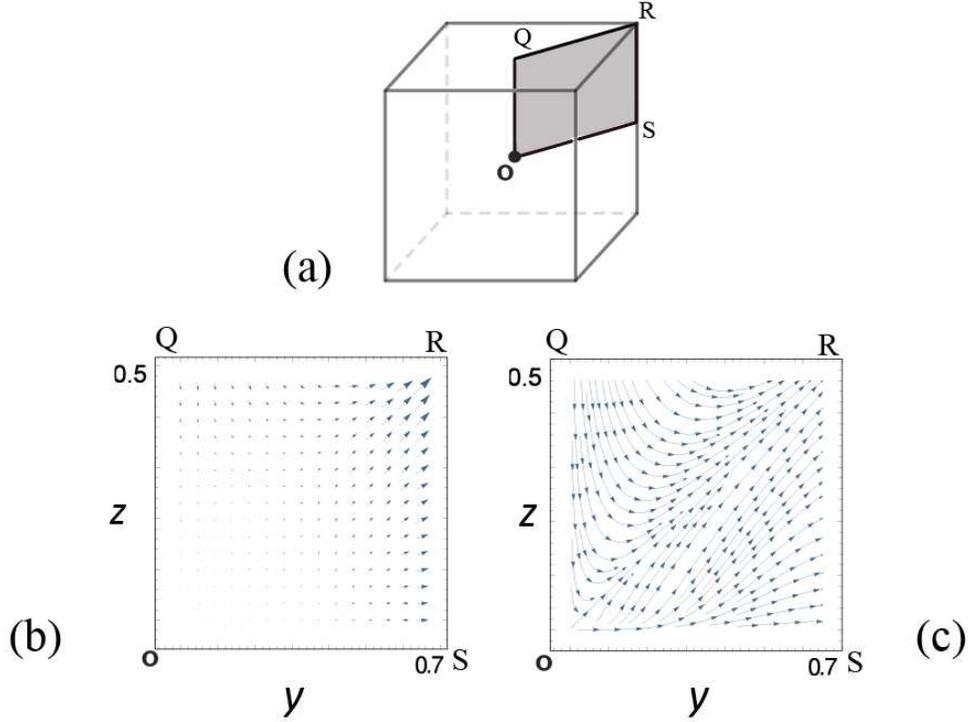}
\caption{ Plots similar to
  Fig.~\ref{fig:vector-streamline-plot-x=0-plane} but now over the
  rectangle~$QRSO$ that lies in a mirror plane that contains the
  diagonal line~${\bf x}(u) = (u,u,u)$. The plots were generated by
  plotting the quantity ${\bf E}(u,u,z)$ over the ranges $0 \le u \le
  0.7$ and $0 \le z \le 0.46$. The vector field ${\bf E}$ now points
  outward from~$O$ to the edge at~$S$ and outward to the corner
  at~$R$. }
\label{fig:vector-streamline-plot-diagonal-plane}
\end{figure}

\begin{figure}
\centering
\includegraphics[width=0.8\textwidth]{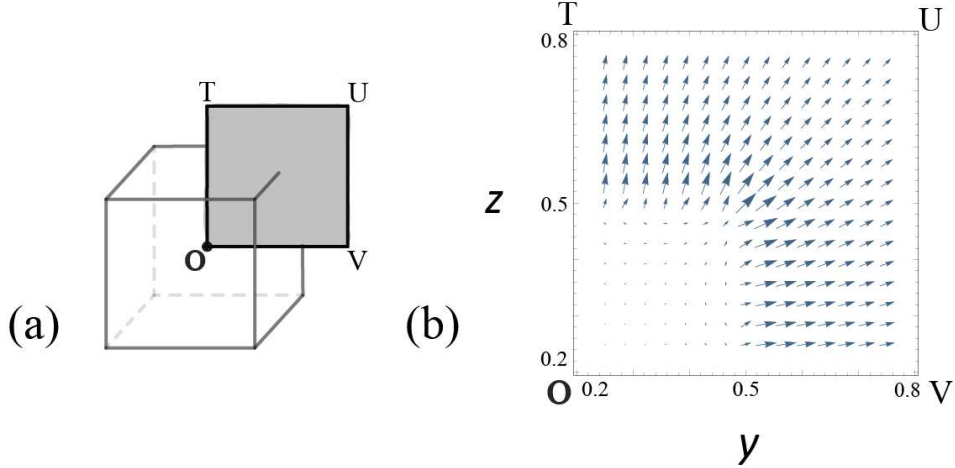}
\caption{(a) Relation of the plotting region, square~$TUVO$, to the
  charged cubic surface. The gray area is the set of points~$x=0$,
  $0.25 \le y < 0.75$, and $0.25 \le z < 0.75$. (b) Vector plot of the
  electric field~${\bf E}(0,y,z)$ over the square~$TUVO$.  We see
  that, in the plane~$x=0$, the external electric field is
  substantially stronger than the internal field, that the electric
  field is particularly large near an edge, and that the external
  electric field is approximately radial.}
\label{fig:vector-plot-E-inside-and-outside-x=0-plane}
\end{figure}

These quantitative observations are reinforced by
figures~\ref{fig:vector-streamline-plot-x=0-plane},
\ref{fig:vector-streamline-plot-diagonal-plane},
and~\ref{fig:vector-plot-E-inside-and-outside-x=0-plane} which show
how the electric field vectors vary over two-dimensional regions that
lie within mirror planes of the charged cubic
surface. Fig.~\ref{fig:vector-streamline-plot-x=0-plane} plots the
electric vector field over a square~$MNPO$ that lies within the mirror
plane~$x=0$.  In agreement with
Sect.~\ref{subsection-application-of-gauss-law} and with
Fig.~\ref{fig:flux-through-front-face}, the vector field plot
Fig.~\ref{fig:vector-streamline-plot-x=0-plane}(a) shows that the
electric field points inwards near midpoints~$M$ and~$P$, and that the
electric field points outwards towards the edge~$N$,

Because the interior electric field magnitude increases as one
approaches the cubic surface and diverges at the edges, the vectors
near the center of the cubic surface are not visible when the vectors
near the surface are displayed with moderate lengths.  This difficulty
can be avoided by using a so-called streamline plot, which consists of
displaying unit electric field vectors~$\hat{\bf E}$ on some fine
regular mesh of spatial points, and then by drawing continuous curves
(the streamlines) that are locally tangent to these unit vectors, see
Fig.~\ref{fig:vector-streamline-plot-x=0-plane}(b). The streamline
plot shows everywhere in the interior how the electric field points
inwards near midpoints of faces and then gradually changes orientation
to point outwards towards the edge at point~$N$.

Finally, Fig.~\ref{fig:vector-plot-E-inside-and-outside-x=0-plane}
shows a vector plot of the electric field over a square region $TUVO$
in the mirror plane~$x=0$ that includes the field external to the
charged cubic surface. As was already shown in
Fig.~\ref{fig:E-V-along-symmetry-lines}(a), the magnitude of the electric
field in the plane~$x=0$ is substantially larger just outside the
cubic surface than anywhere inside so, on the scale of this plot such
that the vectors just outside the surface are of modest size, the
interior electric field vectors are barely visible. The vector plot
confirms the discussion of
Sect.~\ref{subsection-only-interior-field-interesting} in that the
external electric field is qualitatively similar to that of a point
charge at the center~$O$ of the cubic surface, vectors point roughly
away from the center (but this is not a radial field) and decreases in
magnitude as one moves further away from the cubic surface.

\subsection{The electric potential~$V$ associated with the charged
  non-conducting cubic surface.}
\label{subsect-properties-of-V}

In this subsection, we briefly discuss some properties of the electric
potential~$V$ associated with the uniformly charged cubic surface,
using the analytical expression for~$V$ given in
Appendix~\ref{appendix:exact-electric-field}.  We conclude that for
the electric field inside the charged cubic surface, plotting the
vector electric field is more insightful than plotting equipotential
surfaces of the scalar potential~$V$. This seems to contradict the
discussions of introductory physics
textbooks\cite{Knight2012,Tipler2007,Young2011}, but most examples
considered in those books involve just one or two point charges, or
one or two conductors of simple geometry such that the electric field
can be mainly understood by looking at equipotential contours in a
single plane.

\begin{figure}
\centering
\includegraphics[width=0.4\textwidth]{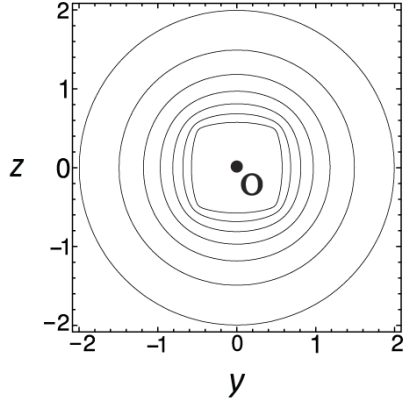}
\caption{Equipotential contours of values~$V=9, 8, 7, 6, 5, 4\cdots,
  3\,\rm V$ of the potential~$V(0,y,z)$ in the plane~$x=0$ external to
  the charged cubic surface of Fig.~\ref{fig:charged-cubic-surface}.
  The contours change smoothly from squarish contours just outside the
  charged cubic surface ($V=9$) to the circular contours of a point
  charge located at the center~$O$ of the cubic surface ($V=3$).}
\label{fig:V-contourplot-external-x=0-plane}
\end{figure}

As was the case for the electric field (see
subsection~\ref{subsection-only-interior-field-interesting}), only the
potential~$V$ inside the charged cubic surface requires understanding
since the potential outside the surface is qualitatively similar to
that of a point charge at the center~$O$ of the surface, as shown in
Fig.~\ref{fig:V-contourplot-external-x=0-plane}.

Panels~(c) and~(d) of Figure~\ref{fig:E-V-along-symmetry-lines}, we
plot the potential~$V$ along the same two symmetry lines as in
panels~(a) and~(b) of the same figure. Because the electric field is
parallel to these symmetry lines for points on these lines (see
subsection~\ref{subsection-E-parallel-to-mirror-planes}), the negative
of the local slope of~$V$ in these plots directly gives the component
of~${\bf E}$ parallel to the symmetry line.  In both cases, the
potential~$V$ asymptotes to the potential~$KQ/d$ of a point charge
at~$O$ (thin black curves in panels~(c) and~(d)) at distances that are
close to the surface of the cube.

From Fig.~\ref{fig:E-V-along-symmetry-lines}(c), we see that the
potential inside the cubic surface is nonuniform and varies over the
modest range $9.6 < V < 9.8$, but this modest variation in value is
misleading since it is the slope of this curve that determines the
magnitude of the electric field. The flat central portion of the curve
near~$x=0$ implies a zero slope and so small electric field component,
which is consistent with the central region of
Fig.~\ref{fig:E-V-along-symmetry-lines}(c). The discontinuous changes
in~$E_x$ from negative to positive finite values at $x=\pm L/2$ in
Fig.~\ref{fig:E-V-along-symmetry-lines}(a) at~$x= 1/2$ are difficult
to see from Fig.~\ref{fig:E-V-along-symmetry-lines}(a) so plotting the
electric field component provides more insight here than plotting~$V$.

Fig.~\ref{fig:E-V-along-symmetry-lines}(d) shows how the potential
varies along the symmetry line~${\bf x}(u) = (u,u,u)$ that passes
through the diagonally opposite vertices $\pm (1/2,1/2,1/2)$. The
range of~$V$ inside the cubic surface along this line is somewhat
broader than in panel~(a), $7.5 \le V \le 9.2$. The potential in the
middle region has again an approximately zero slope, consistent with
the small values of~${\bf E}$ near the center of the cubic
surface. Barely visible at the coordinate values $u = \pm 1/2$ is the
fact that the slope of~$V$ becomes vertical, corresponding to the
logarithmic divergence to infinity of the electric field magnitude at
a vertex of the cubic surface (see
Appendix~\ref{subect-log-divergence}).

Panels~\ref{fig:V-surfaceplots-on-two-interior-planes}(a) and
\ref{fig:V-surfaceplots-on-two-interior-planes}(b) are surface plots
of~$V$ for the same symmetry regions as respectively
Fig.~\ref{fig:vector-streamline-plot-x=0-plane}(a) and
Fig.~\ref{fig:vector-streamline-plot-diagonal-plane}(a). The vector
and streamline plots of~${\bf E}$ clearly provide more insight about
the magnitude and direction of the interior electric field than what
is provided from the corresponding surface plots of~$V$. We observe
that, since the electric field is parallel to these symmetry
rectangles at points on these rectangles, the two-dimensional gradient
$(-\partial_y V, -\partial_z V)$ for panel~(a) or $(-\partial_u V,
-\partial_z V)$ for panel~(b) give the full gradient of~$V$. This
means that the direction and magnitude of the electric field is
determined from just these surface plots (which would not be true for
a surface plot on a rectangle that does not lie within a mirror
plane).

For example, in
Fig.~\ref{fig:V-surfaceplots-on-two-interior-planes}(a) and in
Fig.~\ref{fig:V-surfaceplots-on-two-interior-planes}(b), the surface
is approximately flat near the origin, which implies that the gradient
and so the electric field~$-\nabla{V}$ has a small magnitude near the
center of the cubic surface. The subtle change in concavity of~$V$ in
Fig.~\ref{fig:V-surfaceplots-on-two-interior-planes}(a) corresponds to
the electric field pointing inwards along the face center~$P$ and
outwards towards an edge at~$N$. For example, the surface $V(0,y,z)$
has a negative slope from~$y=0$ to $y=1/2$ along the axis~$z=1/2$,
corresponding to the electric field pointing towards the edge at~$N$,
while the same surface~$V(0,y,z)$ has a positive slope from~$z=0$
to~$z=1/2$ along~$y=0$, corresponding to the electric field pointing
inwards from the face's midpoint~$M$. In contrast, in
Fig.~\ref{fig:V-surfaceplots-on-two-interior-planes}(b), $V(u,u,z)$
decreases as~$u$ increases along any constant~$z$, corresponding to
the electric field pointing outwards towards points~$R$ and~$S$ as
shown more clearly in the electric field vector plot
Fig.~\ref{fig:vector-streamline-plot-diagonal-plane}. These surface
plots of the potential simply do not provide as much insight as the
plots of the electric field vectors.

\begin{figure}
\centering
\includegraphics[width=0.8\textwidth]{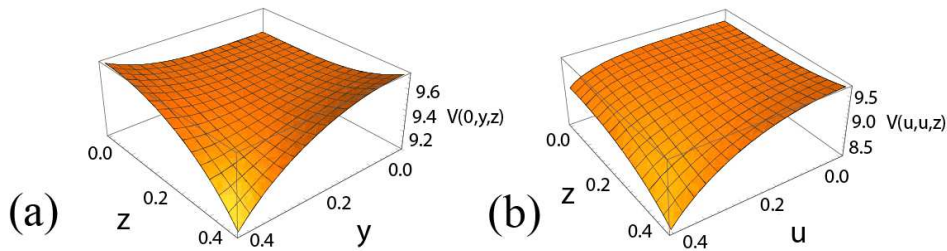}
\caption{ (a) Plot of the potential~$V(0,y,z)$ over the square~$MNOP$
  of Fig.~\ref{fig:vector-streamline-plot-x=0-plane}(a). The negative
  local gradient~$-\nabla{V}$ of this surface, which is difficult to
  determine visually from this plot, corresponds to the directions of
  the electric field in
  Fig.~\ref{fig:vector-streamline-plot-x=0-plane}. (b) Plot of the
  potential~$V(u,u,z)$ over the rectangle~$QRSO$ defined by
  Fig.~\ref{fig:vector-streamline-plot-diagonal-plane}(a). }
\label{fig:V-surfaceplots-on-two-interior-planes}
\end{figure}

\begin{figure}
\centering
\includegraphics[width=0.8\textwidth]{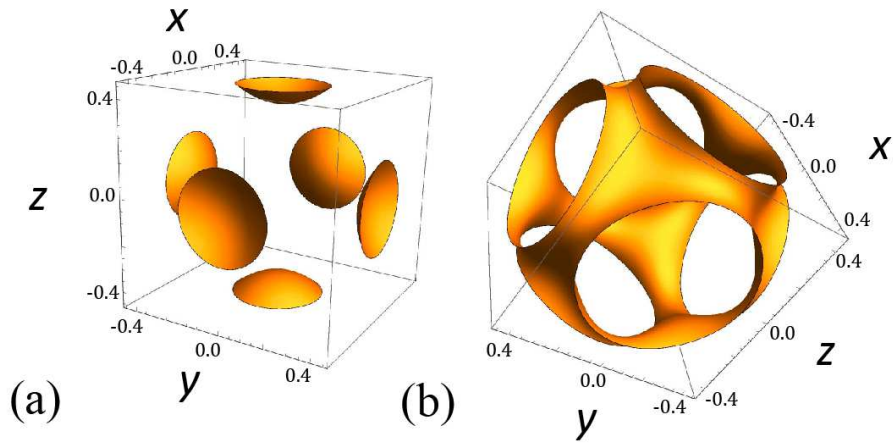}
\caption{ Panels (a) and (b) are equipotential surfaces for
  respectively~$V=9.6$ and~$V=9.4$ inside the cube $|x|, |y|, |z| <
  0.45$ as computed via the Mathematica function {\tt ContourPlot3D}
  using the exact expression Eq.~(\ref{eq-V-cubic-surface-exact}) for
  the potential~$V(x,y,z)$. At each point on these equipotential
  surfaces, the electric field~${\bf E} = -\nabla V$ is perpendicular
  to the surface and points inwards in~(a) and outwards towards edges
  and vertices in~(b). }
\label{fig:two-equipotential-surfaces}
\end{figure}

Finally, Fig.~\ref{fig:two-equipotential-surfaces} shows two
three-dimensional equipotential surfaces for the potential
values~$V=9.6$ and~$V=9.4$ inside the charged cubic surface. (These
were calculated using the Mathematica command \verb|ContourPlot3D|.)
For~$V=9.6$ in Fig.~\ref{fig:two-equipotential-surfaces}(a), the
equipotential surface consists of six disconnected roughly spherical
caps near the midpoint of each face. Since the electric field is
perpendicular to an equipotential surface at any point on that
surface\cite{Knight2012}, Fig.~\ref{fig:two-equipotential-surfaces}(a)
tells us that, near the center of each face, the internal electric
field points inward but over a spread of angles, in agreement with
Fig.~\ref{fig:vector-streamline-plot-x=0-plane}(c) near the
point~$M$. For $V=9.4$ in
Fig.~\ref{fig:two-equipotential-surfaces}(a), the equipotential
surface is geometrically interesting and complex, with the local
normals to this surface being consistent with, but less easy to
understand physically, than the vector plots of
Fig.~\ref{fig:vector-streamline-plot-x=0-plane}(c) and
Fig.~\ref{fig:vector-streamline-plot-diagonal-plane}(c). However,
Fig.~\ref{fig:two-equipotential-surfaces}(b) does help one to
appreciate visually the complexity of the electric field inside the
charged cubic surface.

\section{Conclusions}
\label{sec:conclusions}

In this paper, we have discussed an electrostatics problem concerning
what is the electric field inside and outside of a three-dimensional
symmetric charge distribution consisting of a cubic non-conducting
surface with constant charge density~$\sigma > 0$ (see
Fig.~\ref{fig:charged-cubic-surface}). Although this problem at first
glance seems similar to other problems that students learn about in an
introductory physics course such as the electric field inside a
spherical shell or inside the cavity of a cubic conductor (see
Sect.~\ref{subsect-related-systems-with-zero-field}), the problem has
rather different and interesting features that make it a good choice
for improving a student's conceptual understanding of electrostatics
and for strengthening a student's qualitative and quantitative problem
solving skills. Despite the seeming high symmetry of this charge
distribution, the electric field is not zero inside the cubic surface
and in fact has a rather intricate geometry.

The main purpose of this paper was not so much to describe a new
electrostatics problem but to use this problem as an opportunity to
encourage undergraduate students to use qualitative physical reasoning
at a deeper level than what is traditionally used, even in upper-level
courses in electrodynamics\cite{Griffiths2012}. The authors feel that
this kind of qualitative reasoning, which is used routinely by physics
researchers, is under-emphasized in current physics textbooks. We hope
that this paper will encourage more physics instructors to incorporate
this kind of qualitative reasoning at several points throughout the
semester, by discussing examples that integrate knowledge over
multiple chapters and that apply qualitative, quantitative,
analytical, and numerical approaches to the same problem.

A second contribution of this paper was to emphasize the value of
plotting three-dimensional electric fields over lines or planes that
have certain symmetries with respect to the charge distribution that
produced the electric field (assuming that such symmetries are
present). Nearly all examples of electric fields that are discussed
analytically in undergraduate courses have a simple geometry, e.g.,
they are radial or azimuthal or uniform. The uniformly charged cubic
surface is one of the simplest three-dimensional continuous charge
distributions for which the electric field is intricate and yet can be
mainly understood by qualitative reasoning.

If a course does not have time for students to work through some of
the details of this paper, it would still be worthwhile for the
instructor to mention some of the pedagogical insights that one can
learn from the example of a uniformly charged nonconducting cubic
surface. Some of these insights are the following:
\begin{enumerate}
\item The electric field inside the empty cavity of some static charge
  distribution can be nonzero provided that the charge distribution
  does not have spherical symmetry. Only a conducting surface can
  insure that the electric field is zero everywhere inside the
  interior, no matter what is the symmetry of the conductor.

\item Symmetry is not a binary property that either exists or not
  exists for some object in that one object can be more symmetric than
  another and the differing amounts of symmetry have physical
  consequences.  For example, the surfaces of a sphere and of a cube
  are symmetric but the spherical surface is much more symmetric than
  the cubic surface because a sphere has an infinity of distinct
  mirror planes while a cube has finitely many.  The greater amount of
  symmetry of a spherical surface is enough to force the interior
  electric field to be zero everywhere.

\item The electric field magnitude can diverge to infinity for a
  charge distribution that has a sharp bend like the edges of the
  uniformly charged cubic surface. This can happen even though the
  charge density does not itself diverge in magnitude near an edge or
  vertex, as is the case for a charged conducting surface with a sharp
  bend.

\item For electric fields that vary in magnitude and direction in
  three spatial dimensions, it can be more insightful to visualize the
  electric field rather than equipotential surfaces, contrary to what
  is discussed in many introductory physics
  textbooks\cite{Knight2012,Tipler2007,Young2011}.

\item
  Fig.~\ref{fig:sigma-versus-V-nonconducting-vs-conducting-surfaces}
  makes an important point that should help students learning about
  electrostatics: for non-spherical conductors in electrostatic
  equilibrium, the surface and interior are equipotential but the
  surface charge density is nonuniform. Conversely, for a
  non-spherical surface covered with a uniform charge density, the
  surface and the interior are generally not equipotential.

\item Three-dimensional vector fields like the electric field produced
  by this charged cubic surface are generally difficult to visualize
  and it is rarely insightful to plot the vector field directly (see
  Fig.~\ref{fig:3d-vector-plot-inside-cubic-surface}). Instead,
  several visualization techniques can be used to obtain insight that
  include plotting the field on lines and on rectangles that lie in
  mirror planes, and using streamline plots in addition to vector
  plots to determine the field geometry.

\item Just because an analytical expression is available for
  describing some problem does not automatically imply that the
  expression is scientifically insightful. Although students are
  exposed to this point even during their first few weeks of an
  introductory physics course (it is difficult to understand almost
  any symbolic mathematical expression the first time, say even
  something like the kinematic relation $x= x_0 + v_0 t + (1/2) a t^2$
  for the position of a one-dimensional particle undergoing constant
  acceleration), nearly all mathematical expressions that
  undergraduate physics majors work with tend to be less than one page
  in length, so students get the false impression that all physics
  expressions are this length or shorter. As shown in
  Appendix~\ref{appendix:exact-electric-field}, the potential~$V$ and
  electric field~${\bf E}$ are known explicitly in terms of elementary
  mathematical functions everywhere in space for the charged cubic
  surface, but the expressions are so long so as to be almost useless
  for insight. So even if analytical solutions are available, it is
  important to find ways to understand their properties qualitatively
  as we did in Sect.~\ref{sec:qualitative}, and to find ways to
  visualize their mathematical properties.
\end{enumerate}

Finally, we mention that this problem has many components that make it
well suited for students in a discussion section or flipped class to
work on in groups. Students could work through the subsections of
Sect.~\ref{sec:qualitative} in succession, first learning about how to
use symmetry to constrain the electric field on certain spatial
regions, then using Gauss's law in a qualitative way. They could then
either be given the figures of Sec.~\ref{sec:quantitative} to compare
with their qualitative results or be given a computer code that
evaluates the electric field and potential anywhere in space and be
guided with how to plot these quantities in insightful ways. Students
could also be challenged to explore topics that extend this paper, for
example:
\begin{enumerate}
\item Most of the conclusions obtained by using mirror planes in
  Section~\ref{subsection-E-parallel-to-mirror-planes} can also be
  obtained using the discrete rotational symmetry of various symmetry
  lines, which would be interesting for students to explore.

\item Students could explore a two-dimensional version of this paper:
  consider a square frame that consists of four equal non-conducting
  line segments, each of which that has the same constant linear
  charge density~$\lambda > 0$. Qualitatively and then quantitatively,
  investigate what is the electric field inside and outside the square
  frame?

\item Students could plot the analytical expressions for~${\bf E}$
  and~$V$ given in Appendix~\ref{subsect-derivation-V-and-E} for a
  uniformly charged planar rectangle. Using these plots, they could
  confirm that ${\bf E}$ looks like that of an infinite plane for
  points close to any point that is closer to the rectangle than to an
  edge, that the electric field is not perpendicular to the plane of
  the rectangle as a point of interest approaches an edge, and that
  the electric field asymptotes to that a point charge field far from
  the rectangle. Students could also confirm that electric field
  magnitude~$E$ diverges near edges, which clarifies that the
  divergence of~$E$ near edges of the charged cubic surface arises
  from the properties of each face, not from the cubic geometry.

\item Instead of the charged non-conducting cubic surface, students
  could consider a non-conducting cube filled with a constant volume
  charge density~$\rho$.  They could then investigate qualitatively
  and then quantitatively (say by approximating the density by a
  finite regular grid of point charges) the electric field and
  potential. (An analytical solution for~$V$ is known in this
  case\cite{Ciftja2015}, and a qualitative discussion for a uniform
  cubic mass has been given by Sanny and Smith\cite{Sanny2015}.) How
  do the potential and electrical field differ from those for a sphere
  of constant charge density~$\rho$?

\item Students could investigate qualitatively and then quantitatively
  the internal and external electric field for the the surface of a
  tetrahedron that is covered with a uniform surface charge density.

\end{enumerate}

\appendix 

\section{Analytical expressions for the electric field and potential}
\label{appendix:exact-electric-field}

\subsection{Derivation of the potential and electric field}
\label{subsect-derivation-V-and-E}

In this Appendix, we use superposition and Coulomb's law to obtain an
exact mathematical expression for the potential~$V_{\rm rect}(x,y,z)$
of a uniformly charged rectangle\cite{Hummer1996}. Knowing~$V_{\rm
  rect}$, it is then straightforward to use superposition to obtain an
explicit expression for the potential~$V_C$ anywhere in space of a
uniformly charged cubic surface since this surface consists of six
identical uniformly charged squares, see
Eq.~(\ref{eq-V-cubic-surface-exact}) below. From~$V_C$, one can then
obtain an analytical expression for the electric field anywhere in
space via the relation~${\bf E}=-\nabla V_C$. The authors first
learned about some of these results from an interesting blog of
Michael Trott, who showed how to use Mathematica to calculate and plot
the three-dimensional potential of some charge distributions that have
sharp edges\cite{Trott2012}.

Rather than explain the following details to students, we recommend
that instructors give the students a black-box computer program that
returns as its output the electric field~${\bf E}(x,y,z) =
(E_x,E_y,E_z)$ and the potential~$V(x,y,z)$ at any
point~$(x,y,z)$. This output can then be plotted or studied
numerically.

Consider a rectangle of dimensions $a \times b$ that has a constant
surface charge density $\sigma$. We assume that the rectangle lies in
the $xy$-plane of an $xyz$-Cartesian coordinate system such that the
rectangle's vertices lie at the four points $(x,y,z) = (\pm a/2, \pm
b/2,0)$. The potential~$V_{\rm rect}(x,y,z)$ at some point~$(x,y,z)$
is then given by the following two-dimensional integral:
\begin{equation}
  \label{eq-V-integral-for-rectangle}
  V_{\rm rect}(x,y,z) = \int_{-a/2}^{a/2} \!
  \int_{-b/2}^{b/2}
  { K \sigma \, dx' \, dy' \over
    \left[
      \left( x - x' \right)^2
    + \left( y - y' \right)^2
    +  z^2
    \right]^{1/2} } .
\end{equation}
This integral is a statement of superposition, and is the limit of a
discrete sum over the infinitesimal potentials~$dV = K dq / d$
at~$(x,y,z)$ created by infinitesimal squares of area $dA' = dx'
\times dy'$ and of infinitesimal charges~$dq = \sigma \, dA'$ that are
centered on the point~$(x',y',0)$. By direct evaluation or by using a
symbolic integrator such as those available in Mathematica or Maple,
one finds that this integral has the following value:
\begin{align}
  \label{eq-V-analytical-expression-for-rectangle}
  &  V_{\rm rect}(x,y,z) = \frac{K \sigma}{2} \Bigg( 
    (b-2 y) \log \left( \sqrt{(a-2 x)^2+(b-2 y)^2+4 z^2}+a-2 x \right) \nonumber \\
  & \: \qquad \qquad +  (a-2 x) \log\left(
       \sqrt{(a-2 x)^2+(b-2 y)^2+4 z^2}+b-2 y
     \right) \nonumber \\
  & \: \qquad \qquad - (b-2 y) \log\left(
       \sqrt{(a+2 x)^2+(b-2 y)^2+4 z^2}-a-2 x
     \right) \nonumber \\
  & \: \qquad \qquad +  (a+2 x) \log\left(
    \sqrt{(a+2 x)^2+(b-2 y)^2+4z^2}+b-2 y \right) \nonumber \\
  & \: \qquad \qquad -  (a-2 x) \log\left(
       \sqrt{(a-2 x)^2+(b+2 y)^2+4 z^2}-b-2 y \right) \nonumber \\
  & \: \qquad \qquad - (a+2 x) \log\left(
       \sqrt{(a+2 x)^2+(b+2y)^2+4 z^2}-b-2 y \right) \nonumber \\
  & +  (b+2 y) \bigg(
      \log\left( \sqrt{(a-2 x)^2+(b+2y)^2+4 z^2}+a-2 x \right) \nonumber \\
  & \: \qquad \qquad 
       - \log\left(\sqrt{(a+2 x)^2+(b+2 y)^2+4 z^2}-a-2 x \right) \bigg) \nonumber \\
  &  -2 z \Bigg[
      \tan ^{-1}\left(\frac{(a-2 x) (b-2 y)}{2 z \sqrt{(a-2 x)^2+(b-2 y)^2+4 z^2}}
         \right) \nonumber \\
  & \qquad + \tan^{-1}\left(
       \frac{(a+2 x) (b-2 y)}{2 z \sqrt{(a+2 x)^2+(b-2 y)^2+4
           z^2}}\right) \nonumber \\
  & \qquad + \tan^{-1}\left(
         \frac{(a-2 x) (b+2 y)}{2 z \sqrt{(a-2 x)^2+(b+2 y)^2+4 z^2}}
      \right) \nonumber \\
  & \qquad  + \tan^{-1}\left(
        \frac{(a+2 x) (b+2 y)}{2 z \sqrt{(a+2 x)^2+(b+2 y)^2+4
     z^2}}\right) \Bigg] \; \Bigg) .
\end{align}

The potential~$V_{\rm cube}(x,y,z)$ at any point~$(x,y,z)$ for a
uniformly charged cubic surface centered at the origin with side
length~$L$ is then obtained by first setting the rectangular
lengths~$a=b=L$ and then by adding shifted versions of
Eq.~(\ref{eq-V-analytical-expression-for-rectangle}) like this:
\begin{align}
  V_C(x,y,z) &= 
    \quad   V_{\rm rect}(x,y,z+1/2) + V_{\rm rect}(x,y,z-1/2) \nonumber \\ 
  & \quad + V_{\rm rect}(z,x,y+1/2) + V_{\rm rect}(z,x,y-1/2) \nonumber \\
  & \quad + V_{\rm rect}(z,y,x+1/2) + V_{\rm rect}(z,y,x-1/2) . 
     \label{eq-V-cubic-surface-exact}
\end{align}
Using a symbolic manipulation program like
Mathematica\cite{Mathematica15,Maple15}, one can then obtain an
explicit symbolic expression for the electric field~${\bf E}_{\rm
  cube} = - \nabla{V_{\rm cube}}$ by symbolic partial differentiation
of Eq.~(\ref{eq-V-cubic-surface-exact}) with respect to the
variables~$x$, $y$, and~$z$.  For example, the following brief
Mathematica code defines a function~\verb|ECubicSurface| that returns
a symbolic expression for the electric field vector~${\bf E}(x,y,z)$
at a given point~$(x,y,z)$:
\begin{verbatim}
      ECubicSurface[ x_, y_, z_ ] := Module[
        { x0, y0, z0 } ,
        - Grad[ VCubicSurface[ x0, y0, z0 ] , { x0, y0, z0 } ] 
          /. { x0 -> x, y0 -> y, z0 -> z }
      ] ;
\end{verbatim}
The line \verb|Grad[ VCubicSurface[ x0, y0, z0 ] , { x0, y0, z0 } ]|
takes the symbolic gradient (partial derivatives) of the expression
$V_{\rm cube}(x_0,y_0,z_0)$ with respect to the
vector~$(x_0,y_0,z_0)$. The following line
\verb|/. { x0 -> x, y0 -> y, z0 -> z }| performs a symbolic
substitution, replacing all symbols $x_0, y_0, z_0$ in the expression
for the electric field with respectively the values~$x,y,z$.  The
resulting expression for~${\bf E}$ is several pages long but is
readily evaluated as needed.

\subsection{Logarithmic divergence of the electric field near a
  corner}
\label{subect-log-divergence}

Using these exact expressions and Mathematica, we can show that the
magnitude of the electric field diverges logarithmically as one
approaches any edge or vertex of the charged cubic surface. This means
that, if~$P$ is some point and~$E(P)$ is the magnitude of the electric
field at~$P$, then $E \propto |\log(d)|$ in the limit that the
distance~$d$ of~$P$ to an edge or vertex becomes small ($d \ll L$).
For example, the Mathematica code
\begin{verbatim}
         Series[
           ECubicSurface[ 1/2 - x, 1/2 - x, 1/2 - x ] ,
           { x, 0, 2 } ,
           Assumptions -> ( x > 0 )
         ]
\end{verbatim}
evaluates the Taylor series of ${\bf E}(1/2-x,1/2-x,1/2-x)$ to second
order in the small quantity~$x$ about~$x=0$, which corresponds to the
vertex $(1/2,1/2,1/2)$. Evaluating this code gives the answer
\begin{equation}
  \label{eq-log-divergence-in-E}
  {\bf E} = \left( -2 \log(x) -2.5 - 0.59 x + 3.4x^2 \right) 
            \left( \hat{\bf x} + \hat{\bf y} + \hat{\bf z} \right) .
\end{equation}
Eq.~(\ref{eq-log-divergence-in-E}) says that, as one approaches the
vertex~$(1/2,1/2,1/2)$ along the line $(1/2-x,1/2-x,1/2-x)$, with~$x$
becoming small, the electric field diverges as~$-2\log(x)$.

\subsection{Validation of the symbolic expression using a 
  simple numerical code based on discretization and superposition}
\label{subect-simple-numerical code}

Knowing the exact result
Eq.~(\ref{eq-V-analytical-expression-for-rectangle}) and its gradient
does not automatically imply that one can evaluate it correctly with a
computer program since there are multiple ways that an error can enter
during the process of writing and executing a Mathematica program. To
make sure that our results were correct, we developed an independent
numerical method and then used that method to confirm the correctness
of all the figures in this paper.

\begin{figure}
\centering
\includegraphics[width=0.3\textwidth]{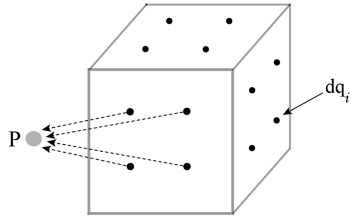}
\caption{The electric field or potential at a given point~$P$ in space
  is obtained by approximating each face of the charged cubic surface
  with a regular square grid of identical point charges and then by
  adding up the electric field or potential due to each point charge
  at~$P$.}
\label{fig:superposition-over-point-charges}
\end{figure}

We did this by using a simple algorithm whose technical details can be
easily understood by freshman physics students, although it can be
challenging to program the algorithm for general charge
densities~$\rho(x,y,z)$. The key idea is illustrated in
Fig.~\ref{fig:superposition-over-point-charges} and the key steps are
summarized here:
\begin{enumerate}
\item the continuous charge distribution on each face of the cubic
  surface is approximated with a square mesh of $N \times N$ identical
  point charges, each of charge $\Delta{Q} = Q/N^2$;

\item at some point~$P=(x,y,z)$ of interest, sum the contributions of
  each of the $6N^2$ point charges to the electric field and potential
  at~$P$, using the elementary expressions $\Delta{\bf E}_i = K
  \Delta{Q}_i \hat{\bf r} / d_i^2$ and $\Delta{V}_i = K \Delta{Q}_i /
  d$. Here~$i$ is some integer label that goes over all the point
  charges~$\Delta{Q}_i$, $d_i$ is the distance between~$P$ and the
  $i$th charge, and $\hat{\bf r}$ is the unit vector pointing from the
  $i$th point charge to~$P$.
\end{enumerate}
We found that a value of~$N \ge =10$ (at least 100~ point charges per
face) gave identical results at the level of the figures when compared
to the exact answer. Some small differences between the exact and
numerical results were found near edges (where the electric field
magnitude diverges) and when some point of interest was close to the
discrete grid of point charges (less than a few multiples of the
spacing between the grid points).

\begin{acknowledgments}
We thank Wenqi Wang for calculating the surface charge density on a
cubic conductor using the COMSOL program and for sharing her results
with us.
\end{acknowledgments}


\bibliographystyle{prsty} 

\bibliography{charged-cube}

\end{document}